\documentclass[%
superscriptaddress,
showpacs,preprintnumbers,
nobibnotes,
amsmath,amssymb,
aps,
prl,
twocolumn
]{revtex4-1}
\usepackage{ulem}
\usepackage{graphicx}
\usepackage{dcolumn}
\usepackage{bm}
\usepackage{multibib}

\begin {document}

\title{Universal relation between instantaneous diffusivity and radius of gyration of proteins in aqueous solution}

\author{Eiji Yamamoto}
\email{eiji.yamamoto@sd.keio.ac.jp}
\affiliation{%
Department of System Design Engineering, Keio University, Yokohama, Kanagawa 223-8522, Japan
}%

\author{Takuma Akimoto}
\affiliation{%
Department of Physics, Tokyo University of Science, Noda, Chiba 278-8510, Japan
}%

\author{Ayori Mitsutake}
\affiliation{%
Department of Physics, Meiji University, Kawasaki, Kanagawa 214-8571, Japan
}%

\author{Ralf Metzler}
\affiliation{%
Institute of Physics \& Astronomy, University of Potsdam, 14476 Potsdam-Golm, Germany
}%



\begin{abstract}
Protein conformational fluctuations are highly complex and exhibit long-term
correlations. Here, molecular dynamics simulations of small proteins demonstrate
that these conformational fluctuations directly affect the protein's instantaneous 
diffusivity $D_I$. We find that the radius of gyration $R_g$ of the proteins
exhibits $1/f$ fluctuations, that are synchronous with the fluctuations of $D_I$.
Our analysis demonstrates the validity of the local Stokes-Einstein type
relation $D_I\propto1/(R_g + R_0)$, where $R_0\sim0.3$ nm is assumed to be a hydration layer around the protein.
From the analysis of different protein types with both strong and weak conformational fluctuations
the validity of the Stokes-Einstein type relation appears to be a general property.
\end{abstract}

\maketitle

Diffusion of colloidal particles in a bulk liquid, known as Brownian motion, is
driven by collisions with the surrounding liquid molecules. Its ensemble-averaged
mean squared displacement (MSD) $\langle\bm{r}(t)^2\rangle=2dDt$ grows linearly
with time, where $d$ is the spatial dimension, $\bm{r}(t)$ the particle position,
and $D$ the diffusion coefficient. In a high-viscous liquid, $D$ of a spherical
particle of radius $R$ follows the classical Stokes-Einstein (SE) relation $D=k_B
T/6\pi\eta R$, where $\eta$ is the viscosity and $k_BT$ thermal energy. In a
coarse-grained view, the radius $R$ of a diffusing particle is typically assumed
to be constant.

The SE-type relation is also valid for the diffusion of proteins, $D\propto1/R_H$, where $R_H$ is the hydrodynamic radius of a protein.
The translational diffusivity of isolated proteins in solution has been predicted
by its size and shape, e.g. molecular weight \cite{YoungCarroadBell1980,
HeNiemeyer2003}, radius of gyration \cite{TynGusek1990, HeNiemeyer2003}, and
interfacial hydration \cite{HalleDavidovic2003}.
Additionally, complex protein-protein interactions are a determinant factor for protein diffusion in macromolecularly crowded liquids \cite{Minton2001,MetzlerJeonCherstvy2016}.
Interestingly, also 2-dimensional lateral diffusion of transmembrane proteins in
protein-crowded membranes follows an SE-type relation
\cite{JavanainenMartinez-SearaMetzlerVattulainen2017},
while in protein-poor membranes the protein diffusivity follows the logarithmic
Saffman-Delbr\"uck law $D \propto {\rm ln}(1/R)$
\cite{WeisNeefVanKramerGregorEnderlein2013}.

Recently, spatial and temporal fluctuations of the local diffusivity of tracer
particles have been reported in heterogeneous media such as supercooled liquids
\cite{YamamotoOnuki1998}, soft materials \cite{WangAnthonyBaeGranick2009,
WangKuoBaeGranick2012}, and biological systems
\cite{SergeBertauxRigneaultMarguet2008,
ManzoTorreno-PinaMassignanLapeyreLewensteinGarcia2015,
YamamotoKalliAkimotoYasuokaSansom2015,
JeonJavanainenMartinez-SearaMetzlerVattulainen2016,
HeSongSuGengAckersonPengTong2016,
WeronBurneckiAkinSoleBalcerekTamkunKrapf2017,
YamamotoAkimotoKalliYasuokaSansom2017,
LampoStylianidouBacklundWigginsSpakowitz2017,CherstvyNagelBetaMetzler2018}.
The measured tracer dynamics exhibits a non-Gaussian distribution of
displacements, anomalous diffusion with a non-linear $t$-dependence of the MSD,
and dynamical heterogeneity. Specifically the local diffusivity fluctuates
significantly with time due to the influence of heterogeneity in the media,
e.g. clustering, intermittent confinement, structure variation, etc. Numerous
theoretical fluctuating-diffusivity models explain specific features of
the non-Gaussianity and anomalous diffusion
\cite{MassignanManzoTorreno-PinaGarcia-ParajoLewensteinLapeyre2014,
ChubynskySlater2014, UneyamaMiyaguchiAkimoto2015,AkimotoYamamoto2016,
MiyaguchiAkimotoYamamoto2016,CherstvyMetzler2016,ChechkinSenoMetzlerSokolov2017,
TyagiCherayil2017,JainSebastian2018,SabriXuKrapfWeiss2020,Hidalgo-SoriaBarkai2020,
SposiniGrebenkovMetzlerOshaninSeno2020,BarkaiBurov2020,
WangSenoSokolovChechkinMetzler2020}.

Interestingly, a fluctuating diffusivity was observed for polymer models in dilute solutions~\cite{Miyaguchi2017}.
However, the precise influence of the temporal change of the observed particle
itself on the diffusivity fluctuations remains unclear.
Protein molecules represent a uniquely suited system to explore the direct connection between
instantaneous conformation and diffusivity. Namely, incessant protein conformational
fluctuations range from small local conformational changes to large and even global
changes in domain motion and in the folding/unfolding dynamics. Since instantaneous
conformations are expected to affect the instantaneous diffusivity of the proteins,
conformational fluctuations may induce a fluctuating diffusivity of proteins.
If true, it is an interesting question to unveil whether a SE-type relation holds
between the locally fluctuating diffusivity and the protein conformations while
the classical SE relation is established only for a static tracer particle.

Here, we report results from extensive all-atom molecular dynamics (MD) simulations
of small proteins isolated in solution to elucidate the effect of protein
conformational fluctuation on the protein diffusivity. Specifically, we show that
the temporal fluctuations of the instantaneous protein diffusivity $D_I$ {\it directly}
depends on the instantaneous radius of gyration $R_g$ by the SE-type relation
$D_I\propto1/(R_g + R_0)$, where $R_0$ is assumed to be a hydration layer around the protein.


\begin{figure}[tb]
\begin{center}
\includegraphics[width=74 mm,bb= 0 0 1021 1530]{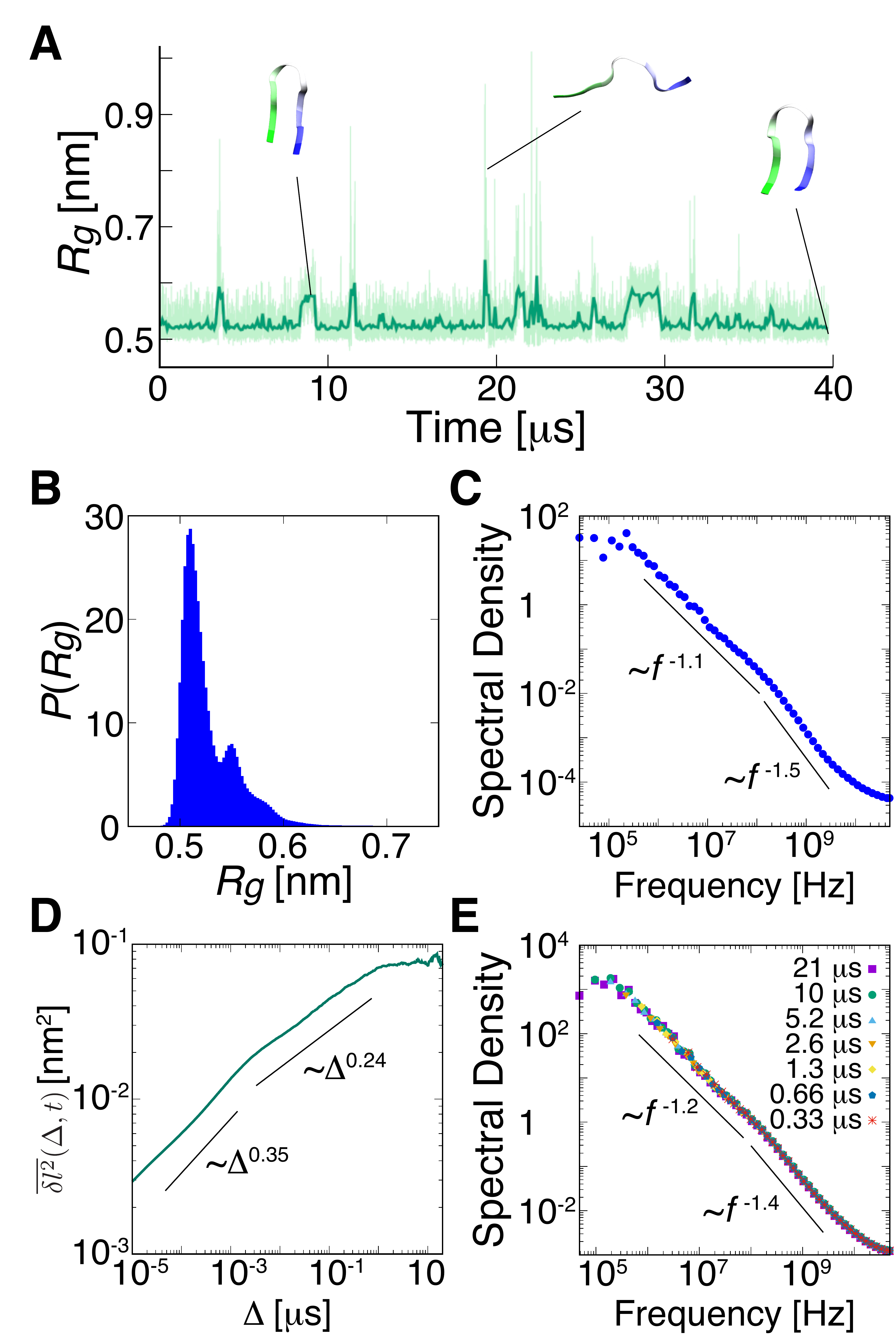}
\caption{Conformational fluctuations of Chignolin at 310~K and 0.1~MPa.
(A)~Time series of the gyration radius $R_g$. Thin and thick lines represent the unsmoothed original values every
1~ns and a smoothed moving average with 100~ns averaging window, respectively.
(B)~Probability density function of $R_g$.
(C)~Ensemble-averaged power spectral density (PSD) of $R_g$
averaged over 5 trajectories of 40~$\mu$s.
Solid lines are shown for reference.
(D)~Ensemble-averaged and time-averaged mean squared protein end-to-end distance for measurement time $t=40$~$\mu$s.
(E)~Ensemble-averaged PSDs of the end-to-end distance.
Different colored symbols represent the PSDs for different measurement times.}
\label{fig_1}
\end{center}
\end{figure}

\begin{figure*}[tb]
\begin{center}
\includegraphics[width=170 mm,bb= 0 0 1903 517]{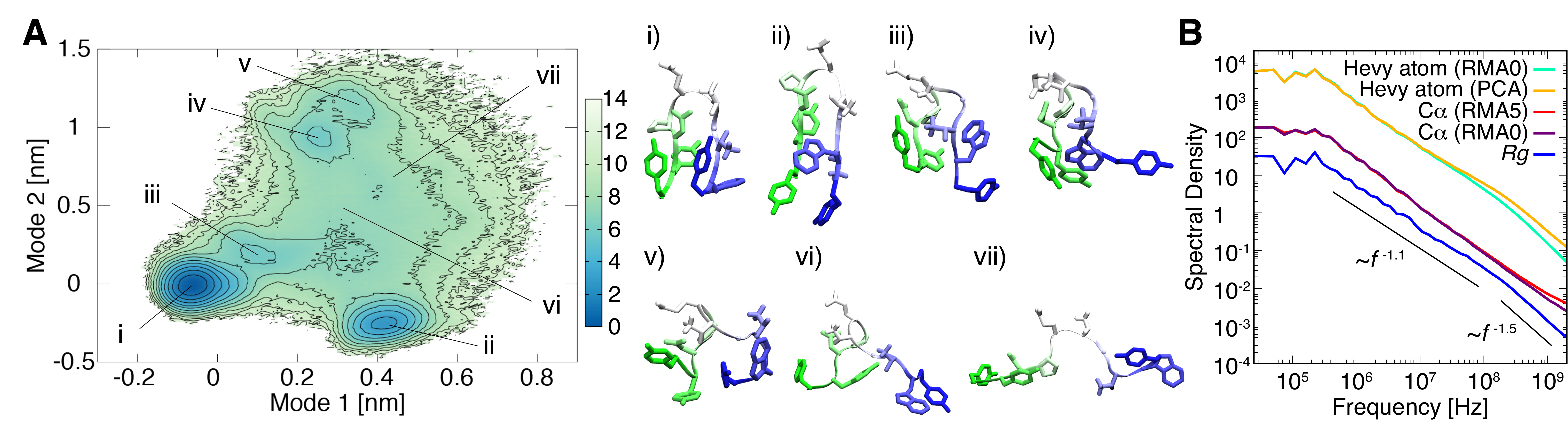}
\caption{Decomposition of the dynamical modes of Chignolin at 310~K and 0.1~MPa.
(A)~Free energy map of relaxation mode~1 v.s. mode~2 obtained by relaxation mode analysis (RMA) using
the coordinates of C$\alpha$ atoms with parameters $t_0=0.5$~ns and $\tau=0.1$~ns.
Snapshots of protein conformations corresponding to the free energy maps: (i) native state, (ii) metastable state, and (iii)-(vii) states 3-7.
Residues 1 to 10 are colored green to blue.
(B)~Ensemble-averaged cumulative PSDs of relaxation modes and principal components. 
RMA and principal component analysis (PCA) were performed using coordinates of heavy atoms or C$\alpha$ atoms.
Parameters for RMA were set as RMA0 ($t_0=0$~ns and $\tau=0.1$~ns) and RMA5 ($t_0=0.5$~ns and $\tau=0.1$~ns).}
\label{fig_rma}
\end{center}
\end{figure*}

\begin{figure}[tb]
\begin{center}
\includegraphics[width=74 mm,bb= 0 0 1170 1145]{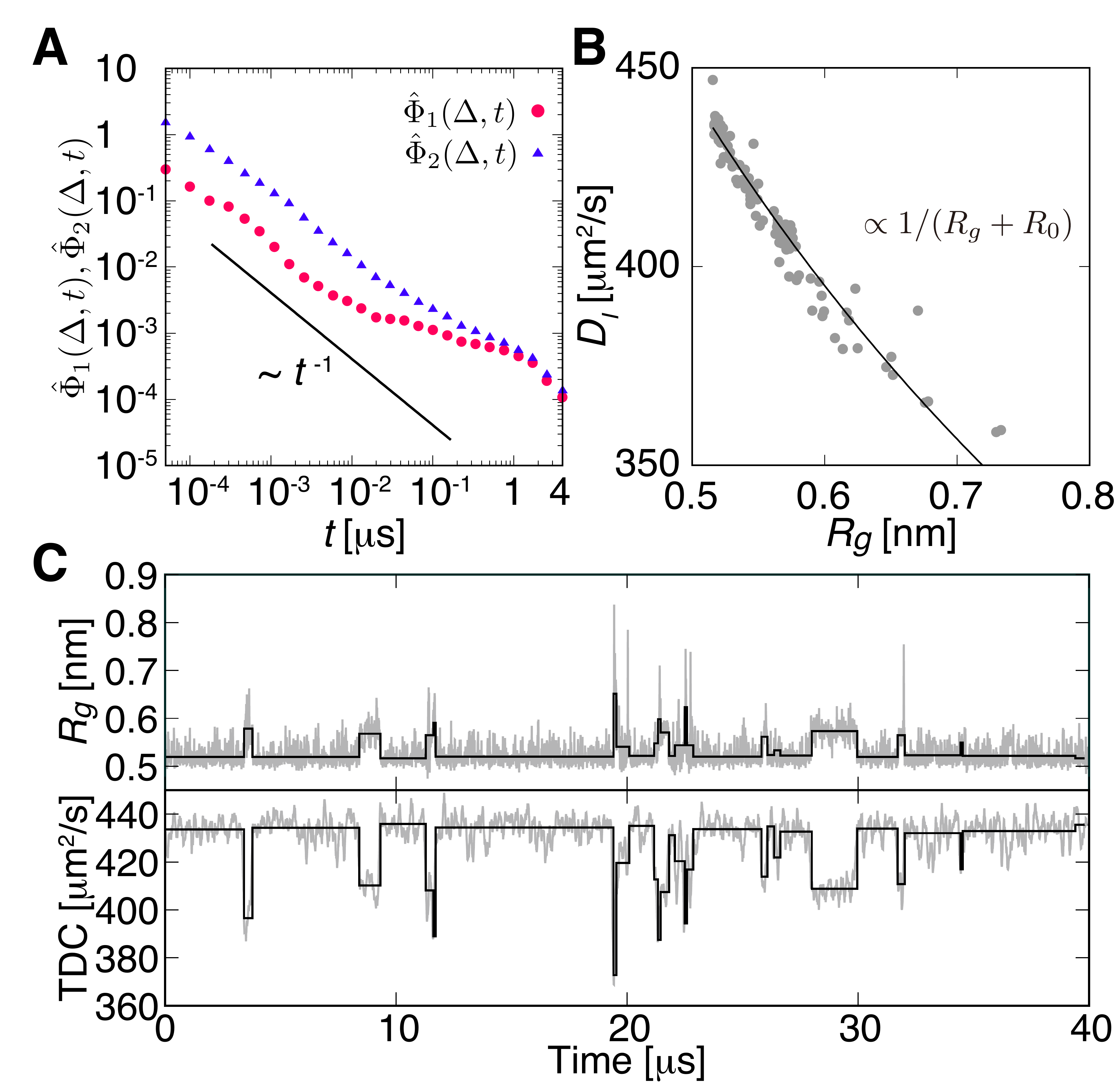}
\caption{Fluctuating diffusivity of Chignolin at 310~K and 0.1~MPa.
(A)~Normalized magnitude $\hat{\Phi}_1(\Delta , t)$ and orientation $\hat{\Phi}_2(\Delta , t)$ correlation functions.
45 divided trajectories were used with a lag time $\Delta = 50$~ps.
(B)~Correlation between the mean $R_g$ and the instantaneous diffusion coefficient $D_I$ in each diffusive state.
(C)~Time series of $R_g$ and temporal diffusion coefficient (TDC).
Thin lines represent unsmoothed original values every 10~ns.
Thick lines represent mean $R_g$ and $D_I$ in each state, where $t=100$~ns and $\Delta=10$~ps were used to obtain the TDC.}
\label{fig_tdc}
\end{center}
\end{figure}

\begin{figure*}[tb]
\begin{center}
\includegraphics[width=150 mm,bb= 0 0 2159 1121]{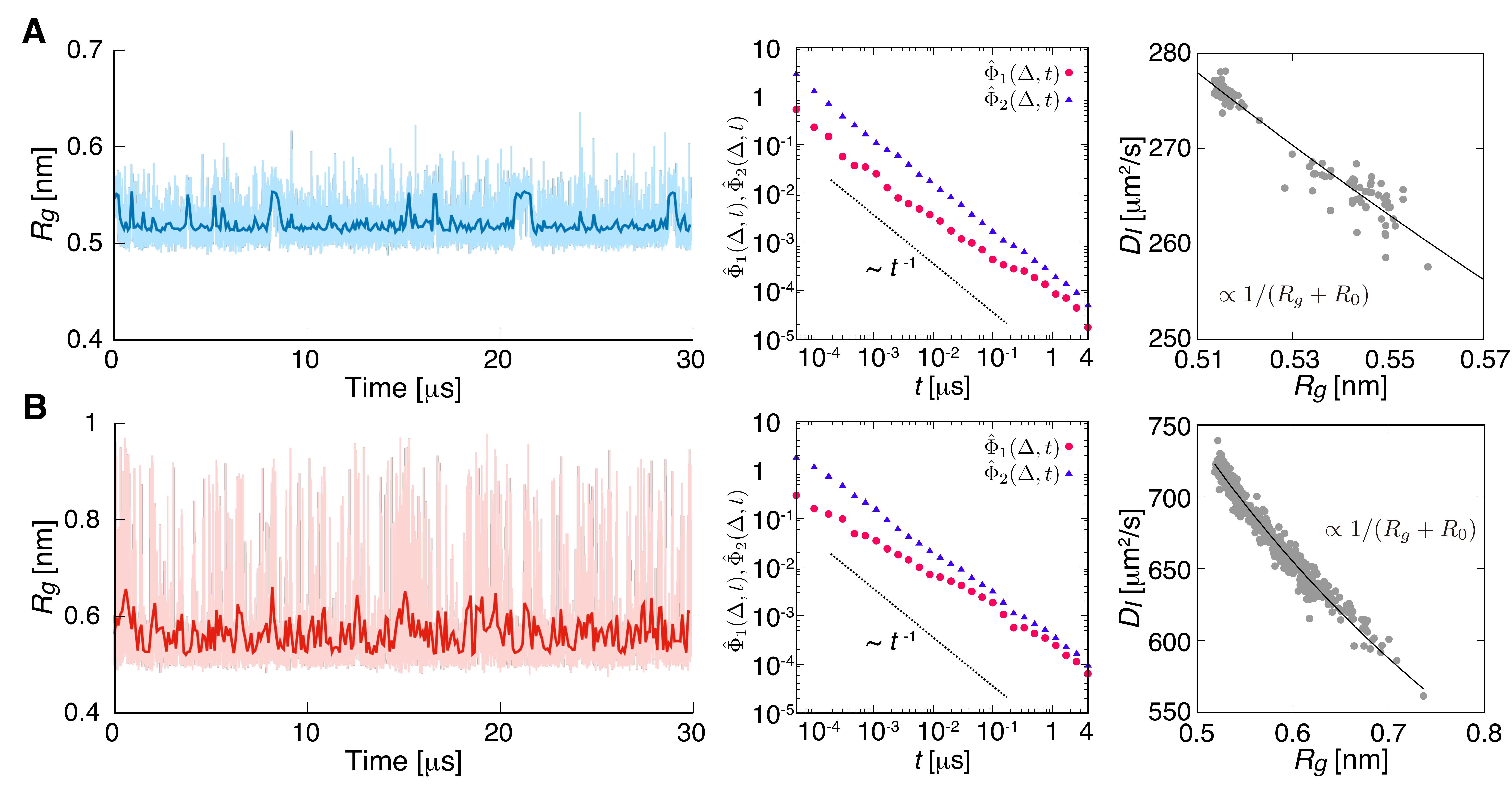}
\caption{Fluctuating diffusivity of Chignolin at different temperature and pressure conditions, (A)~280~K, 0.1~MPa and (B)~400~K, 400~MPa.
(Left) Time series of $R_g$.
Thin and thick lines represent unsmoothed original values every 1~ns and smoothed moving average with 100~ns averaging window, respectively.
(Middle) Normalized magnitude $\hat{\Phi}_1(\Delta, t)$ and orientation $\hat{\Phi}_2(\Delta,t)$ correlation functions.
35 divided trajectories were used with a lag time $\Delta=50$~ps.
(Right) Correlation between mean $R_g$ and $D_I$ in each diffusive state, where $t =100$~ns and $\Delta=10$~ps were used to obtain the TDC.}
\label{fig_diffusion}
\end{center}
\end{figure*}

{\it Conformational fluctuations of Chignolin}--Five independent simulation runs of the protein super Chignolin \cite{HondaAkibaKatoSawadaSekijimaIshimuraOoishiWatanabeOdaharaHarata2008} were run for 40~$\mu$s (see details in SI \cite{support}). To evaluate the conformational fluctuations of Chignolin, the radius of gyration, $R_g^2 = N^{-1} \sum^N_{i=1} \left( \mbox{\boldmath $r$}_i - \mbox{\boldmath $r$}_g \right)^2 $, was calculated, where $N$ is the number of amino acid residues, $\mbox{\boldmath $r$}_i$ and $\mbox{\boldmath $r$}_g$ are the center of mass positions of the $i$th residue and the protein, respectively.
A time series of $R_g$ is shown in Fig.~\ref{fig_1}A. The lower values of $R_g$
corresponds to the folded conformations, while the higher value corresponds to the
unfolded conformations. The probability density function of $R_g$ shows two peaks at
0.51 and 0.55, which correspond to the native state and metastable (misfolded) state,
respectively (see Fig.~\ref{fig_1}B).
Several metastable structures were observed in this simulation of super Chignolin at room temperature (Fig.~\ref{fig_rma}).

Fluctuations of the protein conformations are known to show long-term correlations
\cite{IbenBraunsteinDosterFrauenfelderHongJohnsonLuckOrmosSchulteSteinbachXieYoung1989,
TakanoTakahashiNagayama1998,YangLuoKarnchanaphanurachLouieRechCovaXunXie2003,
YamamotoAkimotoHiranoYasuiYasuoka2014}. Chignolin undergoes a folding and unfolding
transition on a time scale of microseconds. To elucidate the correlations of the
conformational fluctuations, the ensemble-averaged power spectral density (PSD) of
$R_g$ was calculated (Fig.~\ref{fig_1}C and Fig.~S1). The PSD exhibits $1/f$ noise with a power-law exponent of $-1.5$ at high frequencies and $-1.1$ at low frequencies, the transition frequency is $2\times10^8$~Hz.
Below a frequency of $\sim10^6$~Hz, the PSD assumes a plateau, which implies stationarity of the process.
The $1/f$ behavior of the PSD is observed for other small proteins, such as Villin and WW domain of Pin1, whose sizes are about
three times larger than Chignolin, with different power-law exponent (Fig.~S2).

The observed PSD transition frequencies correspond to the time scale of
conformational protein fluctuations. Indeed the time-averaged mean squared
end-to-end distance $\delta l^2$ of Chignolin exhibits a sublinear increase
with two transition points at $\sim1$~ns and $\sim1$~$\mu$s (see Fig.~1D,
details in SI). These transition times are of the same order as those of the
PSD of $R_g$. The PSDs of the end-to-end distance for different measurement
times clearly shows $1/f$ noise similar to that of $R_g$ (Fig.~1E). The
consistency of the PSDs for different measurement times implies absence of
aging \cite{NiemannKantzBarkai2013,SadeghBarkaiKrapf2014,LeibovichBarkai2015}
(see also Fig.~S3). For Chignolin, we clearly see the relaxation of the
conformational fluctuations (plateau in the PSD).

To dissect the dynamical modes of the protein, a relaxation mode analysis (RMA)
\cite{TakanoMiyashita1995,HiraoKosekiTakano1997,MitsutakeIijimaTakano2011,
MitsutakeTakano2015} was performed (see Fig.~\ref{fig_rma} and Figs.~S4-S8).
The free energy maps of relaxation modes (RMs) clearly identify the native state,
metastable state, and other states including unfolded conformations. The slowest
Mode~1 corresponds to a transition between the native and metastable states. The
transition between the native and intermediate states are extracted to the second
slowest Mode~2.
To reveal the origin of the transitions in the PSD of $R_g$, cumulative PSDs summed over 24 individual PSDs of each RM are shown in Fig.~\ref{fig_rma}B.
The cumulative PSD of RMs shows a similar decay as the PSD of $R_g$.
Note that the power-law scaling exponent of the cumulative PSDs converges from $-2$ to $-1.1$ (see Fig.~S7).
This is because the individual PSDs of each RM are expected to exhibit a Brownian noise ($\propto 1/f^2$) due to its exponential relaxation, and the crossover frequency, where the PSD assumes a plateau, corresponds to the relaxation time of its exponential relaxation (Figs.~S4 and S5).
Interestingly, while the cumulative PSD using only the C$\alpha$ atoms does not show the crossover of the power law exponents between $-1.1$ and
$-1.5$ at the transition frequency of $2\times10^8$~Hz, the cumulative PSD using all heavy atoms does show the crossover,
i.e. the crossover at high frequencies originates from the conformational relaxation of side chains.
In addition, the slowest RM of the crossover between the native and metastable states
is related to the crossover frequency where the PSD of $R_g$ assumes a plateau.

{\it Fluctuating diffusivity of Chignolin}--To evaluate the diffusive dynamics of Chignolin in solution,
we calculated the time-averaged MSDs,
\begin{equation}
\overline{\delta {\bm r}^2 }(\Delta , t) = \frac{1}{t-\Delta} \int_0^{t - \Delta } \delta {\bm r}^2 (\Delta, t') dt',
\end{equation}
where $\Delta$ is a lag time, $t$ is the measurement time, and $\delta{\bm r}(\Delta , t') = {\bm r} (t' + \Delta )- {\bm r}(t')$ is the displacement vector of the center of mass position of the protein.
Some scatter was observed where $\Delta$ becomes comparable to $t$ (Fig.~S9).
To examine the fluctuations of the diffusivity, we calculated the magnitude and orientation correlation functions of the diffusivity \cite{Miyaguchi2017, support}.
The magnitude correlation is defined by
\begin{equation}
\Phi_1(\Delta , t) = \langle | \overline{\delta \mbox{\boldmath $r$}^2} (\Delta , t) |^2 \rangle - \langle \overline{\delta \mbox{\boldmath $r$}^2} (\Delta , t) \rangle ^2,
\label{phi_1}
\end{equation}
and the dimensionless form $\hat{\Phi}_1(\Delta , t)$ yields from division by $\langle \overline{\delta {\bm r}^2} (\Delta , t) \rangle ^2$.
$\Phi_1(\Delta , t)$ is equivalent to the ergodicity breaking parameter \cite{HeBurovMetzlerBarkai2008, UneyamaMiyaguchiAkimoto2015, MiyaguchiAkimotoYamamoto2016}.
In the case of ergodic diffusion, e.g. Brownian motion, this parameter converges to 0 with a power-law decay $\propto t^{-1}$.
However, in the case of non-ergodic diffusion \cite{MetzlerJeonCherstvyBarkai2014}, e.g., continuous-time random walks \cite{HeBurovMetzlerBarkai2008, MiyaguchiAkimoto2011a, MiyaguchiAkimoto2011} and annealed transit time models
\cite{AkimotoYamamoto2016a}, the magnitude correlation converges to a non-zero
value for all $\Delta\ll t$ as $t\rightarrow\infty$.
The magnitude correlation function $\hat{\Phi}_1(\Delta,t)$ of Chignolin shows a slow decay with scaling exponent below $-1$, in the time region $t\sim10^{-2}$--$1$~$\mu$s (Fig.~\ref{fig_tdc}A).
This implies that the instantaneous diffusivity may fluctuate intrinsically on the
corresponding time scales. Note that the power-law decay of $-1$ at shorter and longer timescales means that the effect of fluctuating diffusivity can be ignored on these timescales.
The orientation correlation is defined by 
\begin{eqnarray}
\Phi_2(\Delta , t) &=& \langle \overline{\delta \mbox{\boldmath $r$} \delta \mbox{\boldmath $r$}} (\Delta , t) : \overline{\delta \mbox{\boldmath $r$} \delta \mbox{\boldmath $r$}} (\Delta , t) \rangle \nonumber \\
& & - \langle \overline{\delta \mbox{\boldmath $r$} \delta \mbox{\boldmath $r$}} (\Delta , t) \rangle : \langle \overline{\delta \mbox{\boldmath $r$} \delta \mbox{\boldmath $r$}} (\Delta , t) \rangle ,
\end{eqnarray}
where $\overline{\delta{\bm r} \delta{\bm r}}(\Delta , t)$ is a time-averaged MSD tensor \cite{support}, a double dot $:$ is defined by $\mbox{\boldmath $A$} : \mbox{\boldmath $B$} = \sum_{ij} A_{ij}B_{ij}$, and the dimensionless form $\hat{\Phi}_2(\Delta , t)$ yields from division by $\langle \overline{\delta \mbox{\boldmath $r$} \delta \mbox{\boldmath $r$}} (\Delta , t) \rangle : \langle \overline{\delta \mbox{\boldmath $r$} \delta \mbox{\boldmath $r$}} (\Delta , t) \rangle$.
$\hat{\Phi}_2(\Delta,t)$ also shows a slow decay in the time region $t\sim10^{-1}$--$1$~$\mu$s,
i.e. orientational diffusion of the protein fluctuates intrinsically.

Both correlators $\Phi_1(\Delta,t)$ and $\Phi_2(\Delta,t)$ of Chignolin show a
crossover at time $\tau_c\sim1$~$\rm \mu$s, corresponding to the lower crossover
frequency in the PSD of $R_g$ ($\sim 10^6$~Hz). Interestingly, the decays of
$\Phi_1(\Delta,t)$ and $\Phi_2(\Delta,t)$ are similar to those of the flexible
polymer model in dilute solutions, the Zimm model~\cite{Miyaguchi2017}, incorporating
hydrodynamic interactions between monomers (beads) of the polymer \cite{Zimm1956,
ErmakMcCammon1978}. In the Zimm model the correlation function $\langle1/(R_g(t) R_g(0))\rangle$ determines the magnitude of the diffusivity fluctuations \cite{Miyaguchi2017}, and the relaxation time is proportional to the solvent viscosity.
Note that water molecules around biomolecules are known to exhibit subdiffusion~\cite{YamamotoAkimotoYasuiYasuoka2014, TanLiangXuMamontovLiXingHong2018, KrapfMetzler2019}.
Thus, the hydrodynamics interaction within the protein could be more complicated than that of the Zimm model.

To see a direct evidence that the instantaneous diffusivity intrinsically fluctuates with time, we obtained the temporal diffusion coefficient (TDC) at time $t^*$,
\begin{equation}
D(t^*)=\frac{1}{2d\Delta( t -\Delta)}\int_{t^*}^{t^*+ t -\Delta}[\mbox{\boldmath $r$}(t'+\Delta)-\mbox{\boldmath $r$}(t')]^2dt'.
\end{equation}
From the TDC, the transition times of the instantaneous diffusivity $D_I$ were estimated with a statistical test~\cite{AkimotoYamamoto2017, support} (Fig.~S10).
Note that $D_I$ is assumed to be constant between the transition times.
The time series of $D_I$ and mean $R_g$ in each diffusive state fluctuate synchronously.
In particular, $D_I$ decreases when the mean $R_g$ increases (Fig.~\ref{fig_tdc}C). 
A clear relation $D_I\propto1/(R_g + R_0)$ can be seen in Fig.~\ref{fig_tdc}B.
Here, we assume the hydrodynamics radius of the protein is $R_H = R_g + R_0$ with $R_0 = 0.3$ nm, where we interpret the $R_0$ as the hydration layer around the protein.
Note that polymers in the Zimm model with longer chains, that form approximately spherical coils with a radius $R_g$, follow the SE-type relation $D \propto 1/R_g$, i.e. our form when $R_g \gg R_0$~\cite{Miyaguchi2017, DoiEdwards1988}.

The universal nature of the relation between $D_I$ and $R_g$ is underlined by MD simulations of Chignolin under two different temperature and pressure conditions (Fig.~\ref{fig_diffusion}).
At 280~K and 0.1~MPa, where the protein conformation changes little, $R_g$ shows small fluctuations around $R_g=0.51$ to 0.52, but still $R_g$ exhibits $1/f$ noise (Fig.~S10), and the crossover frequency $\sim10^6$~Hz corresponds to the crossover time $\sim
1$~$\rm \mu$s of $\hat{\Phi}_1(\Delta,t)$.
At 400~K and 400~MPa, where the protein exhibits frequent folding and unfolding, $R_g$ shows significant fluctuations on a range of 0.5 to 1.
Now, the crossover time of $\hat{\Phi}_1(\Delta,t)$ is shorter, $\sim0.2$~$\rm \mu$s, which is related to the crossover frequency of the PSD of $R_g$ at ~$5 \times10^6$~Hz (Fig.~S11).
Notably, at both conditions the relation $D_I\propto1/(R_g + R_0)$ was observed with $R_0 = 0.2$ nm (280 K, 0.1 MPa) and $R_0 = 0.3$ nm (400 K, 400 MPa).

{\it Conclusion}--Our study reveals a direct relation between the size fluctuations of proteins, encoded by the time dependence of the gyration radius $R_g$, and their instantaneous diffusivity $D_I$.
Specifically, we uncovered the universal relationship $D_I\propto1/(R_g + R_0)$, representing a time-local SE-type relation.
We also demonstrated that the relaxation of the $R_g$ dynamics is directly related to the conformational transitions in the protein energy landscape.
Both features were studied for the protein Chignolin at different temperature and pressure conditions, as well as for Villin and the WW domain of Pin1 (see Fig.~S12).
In particular, this analysis showed that the SE-type relation holds for both proteins with large and negligible $R_g$-fluctuations.
Note that the prefactors of the scaling $D_I = A / (R_g + R_0)$ for all proteins investigated here are the same order of magnitude of $k_BT/ 6\pi\eta$, and $D_I$ is proportional to $T/\eta$ (Fig. S13).
The relatively small proteins analyzed here exhibit a crossover to stationary dynamics.
We speculate that the instantaneous relationship $D_I\propto1/(R_g + R_0)$ will also hold for larger proteins with more complex dynamics \cite{HuHongSmithNeusiusChengSmith2016} (see also Fig.~S14) and pronounced aging behavior \cite{KrapfMetzler2019}, but this remains to be shown in supercomputing studies.
Such a universal relation would be particularly interesting, as it shows that $D_I$ for even highly unspherical proteins can be sufficiently characterized simply by $R_g$.

Our results provide a microscopic physical rationale for randomly fluctuating diffusivities as encoded in a range of recent modeling approaches. While here we focused on the internal protein dynamics, we speculate that the same SE-type relation will hold for proteins and other tracers moving in complex environments such as biological cells. There on top of potential interactions with the cytoskeleton, tracers are typically not fully inert and may thus accumulate foreign molecules on their surface, leading to time-random instantaneous $R_g$ and thus $D_I$~\cite{EtocBalloulVicarioNormannoLiseSittnerPiehlerDahanCoppey2018}.
Moreover, ongoing multimerization typical for many regulatory proteins may further randomize the tracers' $D_I$ \cite{Hidalgo-SoriaBarkai2020}. This also prompts the question whether similar $R_g$-$D_I$ relations will hold for tracers showing anomalous diffusion \cite{EtocBalloulVicarioNormannoLiseSittnerPiehlerDahanCoppey2018}.

\begin{acknowledgments}
We thank Dr. Takashi Uneyama and Dr. Tomoshige Miyaguchi for fruitful discussion.
This work was supported by Grant for Basic Science Research Projects from the Sumitomo Foundation and Grant ME1535/7-1 from German Research Foundation (DFG).
A. M. also thanks the JSPS KAKENHI Grant Number JP20H03230 for support. 
R.M. also thanks the Foundation for Polish Science (FNP) for support.
\end{acknowledgments}

\clearpage

\onecolumngrid
{
    \center \bf \large 
    Supplementary Materials for ``Universal relation between instantaneous diffusivity and radius of gyration of proteins in aqueous solution''\vspace*{1cm}\\ 
    \vspace*{0.0cm}
}
\twocolumngrid

\setcounter{figure}{0}

\section*{Methods}
\subsection*{Molecular dynamics simulations}
We performed all-atom molecular dynamics (MD) simulations of super Chignolin (10 amino acid residues) (PDB ID:2RVD~\cite{HondaAkibaKatoSawadaSekijimaIshimuraOoishiWatanabeOdaharaHarata2008}), Villin (PDB ID:2F4K~\cite{KubelkaChiuDaviesEatonHofrichter2006}), and WW domain of Pin1 (PDB ID:1PIN~\cite{RanganathanLuHunterNoel1997}) using Gromacs 5.1~\cite{AbrahamMurtolaSchulzPallSmithHessLindahl2015}.
The size of Chignolin, Villin, and WW domain are 10, 35, and 35 amino acid residues, respectively.
Chignolin was solvated in a cubic box of $\sim$4~nm containing 1,856 water molecules.
For Villin and Pin1, the protein was solvated in a cubic box of $\sim$5~nm containing 3,904 water molecules.
NaCl ions were added to neutralize the systems.
For each simulation system, five independent simulations were performed in which initial atom velocities were randomly generated.
All systems were subjected to steepest-descent energy minimization to remove the initial close contacts, and equilibrated for 1~ns in $NPT$ constant simulations.
And then the production runs with $NVT$ constant were performed in which the average box size was determined from the last 0.9~ns data of the NPT simulations.
A timestep of 2.5~fs was used for all simualtions.
For Chignolin, simulations were performed under three temperature and  pressure conditions; i) five 40~${\rm \mu}$s at 310~K and 0.1~MPa, 
ii) five 30~$\mu$s at 280~K and 0.1~MPa, and iii) five 30~$\mu$s at 400~K and 400~MPa.
Under the low temperature condition, the protein was keeping the same conformation.
Conversely, under high temperature and pressure condition, the protein exhibited frequent folding and unfolding dynamics~\cite{Okumura2012}.
For Villin and Pin1, five 20~${\rm \mu}$s simulations were performed at 310~K and 0.1~MPa for each system.
For the analysis trajectory data was saved every 10~ps, and the first 100~ns were excluded  for the equilibration.

The systems were subject to pressure scaling to 1~bar using a Berendsen barostat~\cite{BerendsenPostmaGunsterenDiNolaHaak1984} with a coupling time of 0.5~ps.
The temperature was controlled using velocity-rescaling method~\cite{BussiZykova-TimanParrinello2009} with a coupling time of 0.1~ps.
The AMBER99SB-ILDN force field~\cite{Lindorff-LarsenPianaPalmoMaragakisKlepeisDrorShaw2010} was used for protein with the TIP3P water model~\cite{JorgensenChandrasekharMaduraImpeyKlein1983}.
The H-bond lengths were constrained to equilibrium lengths using the LINCS algorithm~\cite{HessBekkerBerendsenFraaije1997}.
Van der Waals and Coulombic interactions were cut off at 1.0~nm.
Coulombic interactions were computed using the particle-mesh Ewald method~\cite{EssmannPereraBerkowitzDardenLeePedersen1995}.

\subsection*{Time-averaged mean squared end-to-end distance of protein}
The time-averaged mean squared end-to-end distance of protein is defined as 
\begin{equation}
\overline{\delta l^2 }(\Delta , t) = \frac{1}{t-\Delta} \int_0^{t - \Delta } [ l(t' + \Delta )- l(t')]^2dt',
\end{equation}
where $\Delta$ is a lag time, $t$ is the measurement time, $l(t')$ is the distance between the center of mass positions of the C terminal and N terminal residues at time $t'$.
The $\overline{\delta l^2 }(\Delta , t)$ is ensemble averaged over $N$ different $\overline{\delta l^2 }(\Delta , t)$ obtained from independent trajectories.

The autocorrelation function $C'(\Delta,t)$ of the end-to-end distance of protein is given by
\begin{equation}
C'(\Delta , t) = \frac{1}{t-\Delta} \int_0^{t - \Delta }  \delta l(t') \delta l(t' + \Delta) dt',
\end{equation}
with $\delta l(t') = l(t') - \langle l \rangle$, where $\langle l \rangle$ is the average distance.
The autocorrelation function is ensemble averaged over $N$ different $C'(\Delta , t)$ obtained from independent trajectories, and is normalized as 
\begin{equation}
C(\Delta , t) = \langle C'(\Delta , t) \rangle / \langle C'(0 , t) \rangle .
\end{equation}
The $N$ different independent trajectories were generated from MD trajectories divided with the measurement time $t$, i.e. the number of ensembles $N$ is different depending on $t$.

\subsection*{Relaxation mode analysis}
We performed the relaxation mode analysis (RMA) to decompose the modes of protein dynamics from trajectories~\cite{TakanoMiyashita1995, HiraoKosekiTakano1997, MitsutakeIijimaTakano2011, MitsutakeTakano2015, MitsutakeTakano2018}.
Here, we consider the $3N$-dimensional column vector $\mbox{\boldmath $R$}$ composed of atomic coordinates relative to their average coordinates,
\begin{equation}
\mbox{\boldmath $R$}^{\rm T} = ( \mbox{\boldmath $r$}_1'^{\rm T} , \mbox{\boldmath $r$}_2'^{\rm T}, ... , \mbox{\boldmath $r$}_N'^{\rm T} ) = ( x_1' , y_1', z_1', ... , x_N' , y_N', z_N') 
\end{equation}
with $\mbox{\boldmath $r$}_i' = \mbox{\boldmath $r$}_i - \langle \mbox{\boldmath $r$}_i \rangle$, where $\mbox{\boldmath $r$}_i$ is the coordinate of the $i$th atom, $\langle \mbox{\boldmath $r$}_i \rangle$ is its average coordinate after removing the translational and rotational degrees of freedom, $N$ is the number of atoms in the protein.
The RMA approximately estimates the slow relaxation modes and their relaxation rates by solving the generalized eigenvalue problem of the time correlation matrices of the coordinates,
\begin{equation}
\sum^{3N}_{j=1} C_{i, j} (t_0 + \tau) f_{p, j} = e^{- \lambda_p \tau} \sum^{3N}_{j=1} C_{i, j} (t_0) f_{p, j} ,
\end{equation}
where $C_{i, j}(t)$ is the component of the $3N \times 3N$ symmetric matrix $C(t)$ defined by
\begin{equation}
C_{i, j}(t) = \langle R_i(t) R_j(0) \rangle .
\end{equation}
Here, $t_0$ is the evolution time, $\tau$ is a time interval, $\lambda_p$ is the relaxation rate of the estimated relaxation modes $f_{p, j}$, and $\langle ... \rangle$ is the ensemble average.
The parameter $t_0$ is introduced in order to reduce the relative weight of the faster modes contained in $\mbox{\boldmath $R$}$, and better estimation of the slow relation modes is expected with sufficiently large $t_0$.
Note that the tICA~\cite{NaritomiFuchigami2011} is a special case of the RMA with $t_0=0$.
In the RMA, $3N-6$ relaxation modes are obtained because the translational and rotational degrees of freedom are removed from $\mbox{\boldmath $R$}$.
By multiplying $\mbox{\boldmath $f$}_p^{\rm T}$, the relaxation mode $X_p$ is given by
\begin{equation}
X_p \approx \sum^{3N}_{j=1} e^{-\lambda_p t_0/2} f_{p, j} R_j .
\end{equation}
For more details, see Ref.~\cite{MitsutakeTakano2018}.

\subsection*{Magnitude and orientation correlation functions of the diffusivity}
Magnitude and orientation correlation functions of the diffusivity~\cite{Miyaguchi2017} were calculated as following.
The time-averaged mean squared displacement (TMSD) is defined as 
\begin{equation}
\overline{\delta {\bm r}^2 }(\Delta , t) = \frac{1}{t-\Delta} \int_0^{t - \Delta } \delta {\bm r}^2 (\Delta, t') dt',
\end{equation}
where $\Delta$ is a lag time, $t$ is the measurement time, and the displacement vector $\delta{\bm r}(\Delta , t') = {\bm r} (t' + \Delta )- {\bm r}(t')$ is obtained using the center of mass position $\bm{r}(t')$ of the protein at time $t'$. 
A TMSD tensor is defined as 
\begin{equation}
\overline{\delta{\bm r} \delta{\bm r}}(\Delta , t) = \frac{1}{t-\Delta} \int_0^{t - \Delta } \delta {\bm r}(\Delta, t') \delta {\bm r} (\Delta, t')dt',
\end{equation}
where the integral is taken for each element of the tensor.

Two scalar functions $\Phi_1(\Delta , t)$ and $\Phi_2(\Delta , t)$ derived from the forth-order correlation function of the TMSD tensor,
\begin{eqnarray}
{\bm \Phi}(\Delta,t) &=& {\bm\langle} {\bm[}\overline{\delta{\bm r}\delta{\bm r}}(\Delta , t) - \langle \overline{\delta{\bm r}\delta{\bm r}}(\Delta , t) \rangle {\bm]} \nonumber \\
& & {\bm[}\overline{\delta{\bm r}\delta{\bm r}}(\Delta , t) - \langle \overline{\delta{\bm r}\delta{\bm r}}(\Delta , t) \rangle {\bm]} {\bm\rangle},
\end{eqnarray}
represent the magnitude and orientation correlations, respectively.
The magnitude correlation is defined by
\begin{equation}
\Phi_1(\Delta , t) = \langle | \overline{\delta \mbox{\boldmath $r$}^2} (\Delta , t) |^2 \rangle - \langle \overline{\delta \mbox{\boldmath $r$}^2} (\Delta , t) \rangle ^2,
\label{phi_1}
\end{equation}
and the dimensionless form $\hat{\Phi}_1(\Delta , t)$ yields from division by $\langle \overline{\delta {\bm r}^2} (\Delta , t) \rangle ^2$.

The orientation correlation is defined by 
\begin{eqnarray}
\Phi_2(\Delta , t) &=& \langle \overline{\delta \mbox{\boldmath $r$} \delta \mbox{\boldmath $r$}} (\Delta , t) : \overline{\delta \mbox{\boldmath $r$} \delta \mbox{\boldmath $r$}} (\Delta , t) \rangle \nonumber \\
& & - \langle \overline{\delta \mbox{\boldmath $r$} \delta \mbox{\boldmath $r$}} (\Delta , t) \rangle : \langle \overline{\delta \mbox{\boldmath $r$} \delta \mbox{\boldmath $r$}} (\Delta , t) \rangle ,
\end{eqnarray}
where a double dot $:$ is defined by $\mbox{\boldmath $A$} : \mbox{\boldmath $B$} = \sum_{ij} A_{ij}B_{ij}$, and the dimensionless form $\hat{\Phi}_2(\Delta , t)$ yields from division by $\langle \overline{\delta \mbox{\boldmath $r$} \delta \mbox{\boldmath $r$}} (\Delta , t) \rangle : \langle \overline{\delta \mbox{\boldmath $r$} \delta \mbox{\boldmath $r$}} (\Delta , t) \rangle$.

\subsection*{Detection of transition times of diffusivity}
The transition of the diffusive states were obtained as follows (see Ref.~\cite{AkimotoYamamoto2017}).
The $d$-dimensional temporal diffusion coefficient (TDC) at time $t^*$ is defined by
\begin{equation}
D(t^*)=\frac{1}{2d\Delta( t -\Delta)}\int_{t^*}^{t^*+ t -\Delta}[\mbox{\boldmath $r$}(t'+\Delta)-\mbox{\boldmath $r$}(t')]^2dt'.
\end{equation}
where $\Delta$ and $t$ are parameters.
The lag time $\Delta$ can be set to the minimal time step of the time series if the time step is greater than the characteristic time of the ballistic motion.
The measurement time $t$ is a tuning parameter that must be smaller than the characteristic time of the diffusive state.
Here, the parameters were set as $t=100$~ns and $\Delta=10$~ps.

Next, we detect transitions of diffusive states using a multiple threshold method, where the $w$th threshold is $D_{\rm eff}^w (w = 1, ..., n)$, for the TDC.
The intervals between consecutive thresholds are set to be constant and the length of the intervals depends on the time series of the TDC.
For example, in the case of Chignolin at 310~K and 0.1~MPa, the thresholds were set every 15~${\rm \mu m^2/s}$ (see Fig.~S10A).
Note that if the length of the interval is set to be too small, the statistical test described below fails to correct the transition points. Therefore, one should determine the multiple thresholds adequately.

For each threshold $D_{\rm eff}^w$, the crossing points $c_i$ are defined by the times at which the TDC crosses $D_{\rm eff}^w$, i.e. $D(c_i) < D_{\rm eff}^w $ and $D(c_i +h) > D_{\rm eff}^w $ or $D(c_i) > D_{\rm eff}^w $ and $D(c_i + h) < D_{\rm eff}^w $, satisfying $c_{i+1} - c_i > t$, where $h$ is the time step of the time series.
The crossing points are not exact points representing changes in the diffusive states because different diffusive states coexist in a time window $[t^*, t^* + t - \Delta]$ of $D(t^*)$.
Therefore, the transition time is defined as $t_i \equiv c_i + t/2 $.
The term $t/2 $ is not exact when the threshold is not at the middle of two successive diffusive states.

The transition times obtained above were corrected with a statistical test for obtaining the exact transition points.
The diffusion coefficient of the $i$th diffusive state in the time interval $[t_i, t_{i+1}]$ is given by
 \begin{equation}
\overline{D}_i \equiv \frac{ \int_{t_i}^{t_{i+1}-\Delta}\{\bm{r}(t'+\Delta) - \bm{r}(t')\}^2dt'}{2d\Delta(t_{i+1}-t_i-\Delta)}.
\label{DC_i}
\end{equation}
Since we consider a situation that the measurement time $t$ is sufficiently large ($t/h >30$), fluctuations of $\overline{D}_i$ can be approximated as a Gaussian distribution with the aid of the central limit theorem.
According to a statistical test, the $i$th and $j$th states can be considered as the same state if there exists $D$ such that both the $k=i$ and $k=j$ states satisfy
\begin{equation}
 D - \sigma_k Z \leq \overline{D}_k \leq D + \sigma_k Z,
\label{statistical_test}
\end{equation}
where  $\sigma_{k}^2$ is the variance of the TDC in the time interval $t_{k+1}-t_k$ and the diffusion coefficient $D$, which is given by $\sigma_{k}^2 \equiv \frac{4 ((\overline{D}_k + \overline{D}_{k+1})/2)^2 \Delta }{3(t_{k+1}-t_k)}$, $Z$ is determined by the level of statistical significance, e.g., $Z=1.96$ when the $p$-value is 0.05.
Therefore, the transition times can be corrected if the two successive diffusion states are the same.
We repeated this procedure: Eq.~(\ref{DC_i}) is calculated again after correcting the transition times $t_i$, and the above statistical test is repeated to correct the transition times.

Finally, merging the transition times obtained for each $D_{\rm eff}^w$, we repeat the statistical test to correct the merged transition times.
Figure S10 shows results of detecting transition times using various parameters, $\Delta$ and $t$.
The instantaneous diffusion coefficients $D_I$ were calculated using Eq.~(\ref{DC_i}) with the obtained transition times.
The mean $R_g$ of the $i$th diffusive state in the time interval $[t_i, t_{i+1}]$ was calculated as 
 \begin{equation}
\overline{R_g}_i \equiv \frac{ \int_{t_i}^{t_{i+1}}  R_g(t') dt'}{t_{i+1}-t_i}.
\label{RG_ave}
\end{equation}
We confirmed no parameter dependence on the correlation between the mean $R_g$ and the instantaneous diffusion coefficient $D_I$ in each diffusive state.

\begin{figure*}[h]
\begin{center}
\includegraphics[width=60 mm,bb= 0 0 156 159]{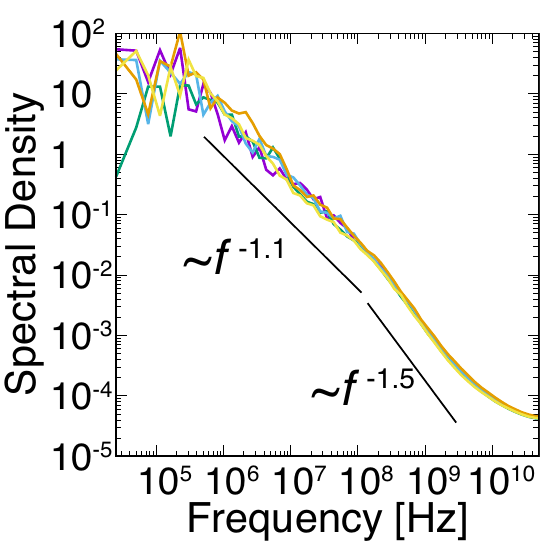}
\caption{Power spectral densities (PSDs) of $R_g$ of Chignolin.
Different colored lines represent the PSDs obtained from five independent simulations.
Solid lines are shown as a reference for power-law decays in higher and lower frequencies.}
\label{fig_radius_gyration_fftw_multi}
\end{center}
\end{figure*}

\begin{figure*}[h]
\begin{center}
\includegraphics[width=120 mm,bb= 0 0 313 167]{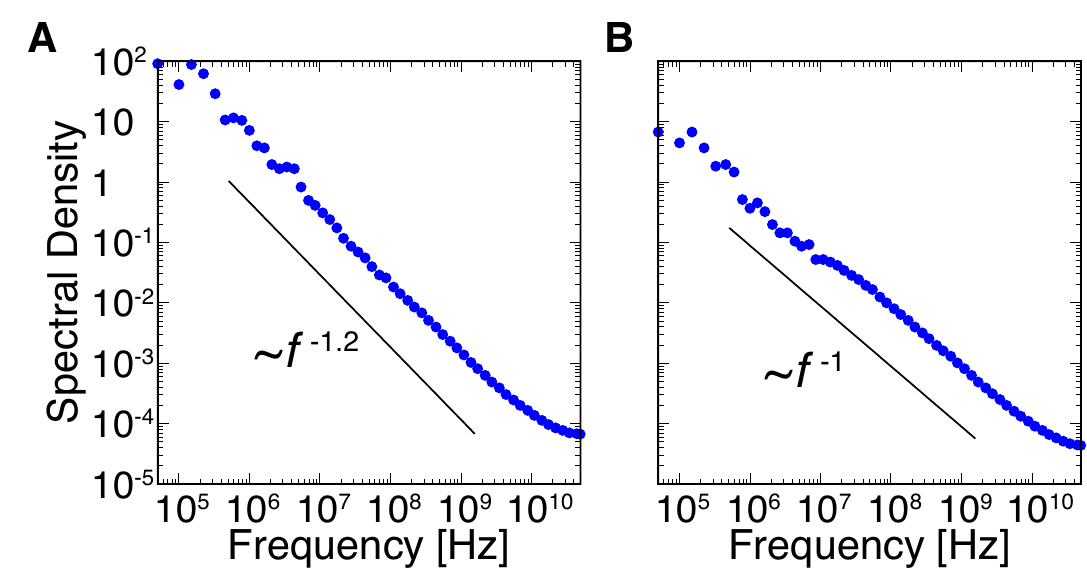}
\caption{Ensemble-averaged PSDs of $R_g$ of (A)~Villin and (B)~WW domain of Pin1.
Solid line is shown as a reference for a power-law exponent.
Because the relaxation time of conformational fluctuation is longer than this time scale, plateau does not appear in the PSD at low frequency.}
\label{fig_FS_radius_gyration_fftw_villin_pin1}
\end{center}
\end{figure*}

\begin{figure*}[h]
\begin{center}
\includegraphics[width=130 mm,bb= 0 0 1484 1061]{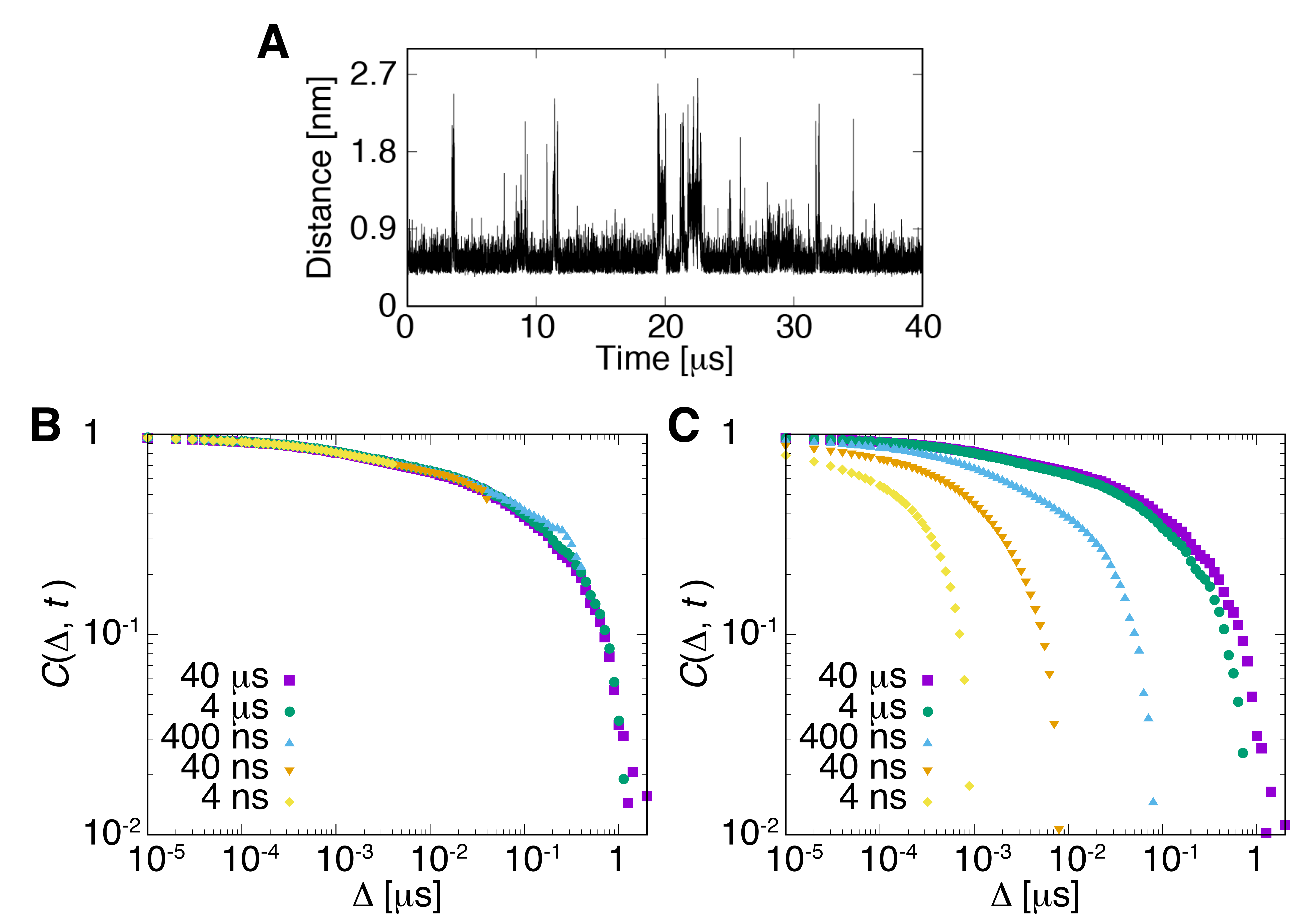}
\caption{Fluctuation of the end-to-end distance of Chignolin.
(A)~A time series of the end-to-end distance of Chignolin.
The coordinates used for the analysis is the same as those in Fig.~1A.
(B)~Normalized autocorrelation functions (NAFs) of the end-to-end distance, where $\langle l \rangle$ was the average over all five 40~$\mu$s simulations.
The different colored symbols represent the NAFs with different measurement time $t$.
The autocorrelation function was ensemble averaged over $N$ different independent trajectories, which were generated from MD trajectories divided with the measurement time $t$.
This result means no aging behavior of the NAF.
According to Wiener--Khinchin theorem, the autocorrelation function and PSD are related.
The no aging behavior is consistent with the PSD (see Fig.~1E (main text)).
(C)~NAFs of the end-to-end distance, where $\langle l \rangle$ was averaged over each independent trajectory, i.e. $\langle l \rangle = \int_0^t l(t') dt' / t $.
In this case, although aging like behavior is observed in the NAFs, this is an analytical error for using the wrong definition of $\langle l \rangle$.}
\label{fig_FS_distance_fluctuation}
\end{center}
\end{figure*}

\begin{figure*}[h]
\begin{center}
\includegraphics[width=140 mm,bb= 0 0 423 562]{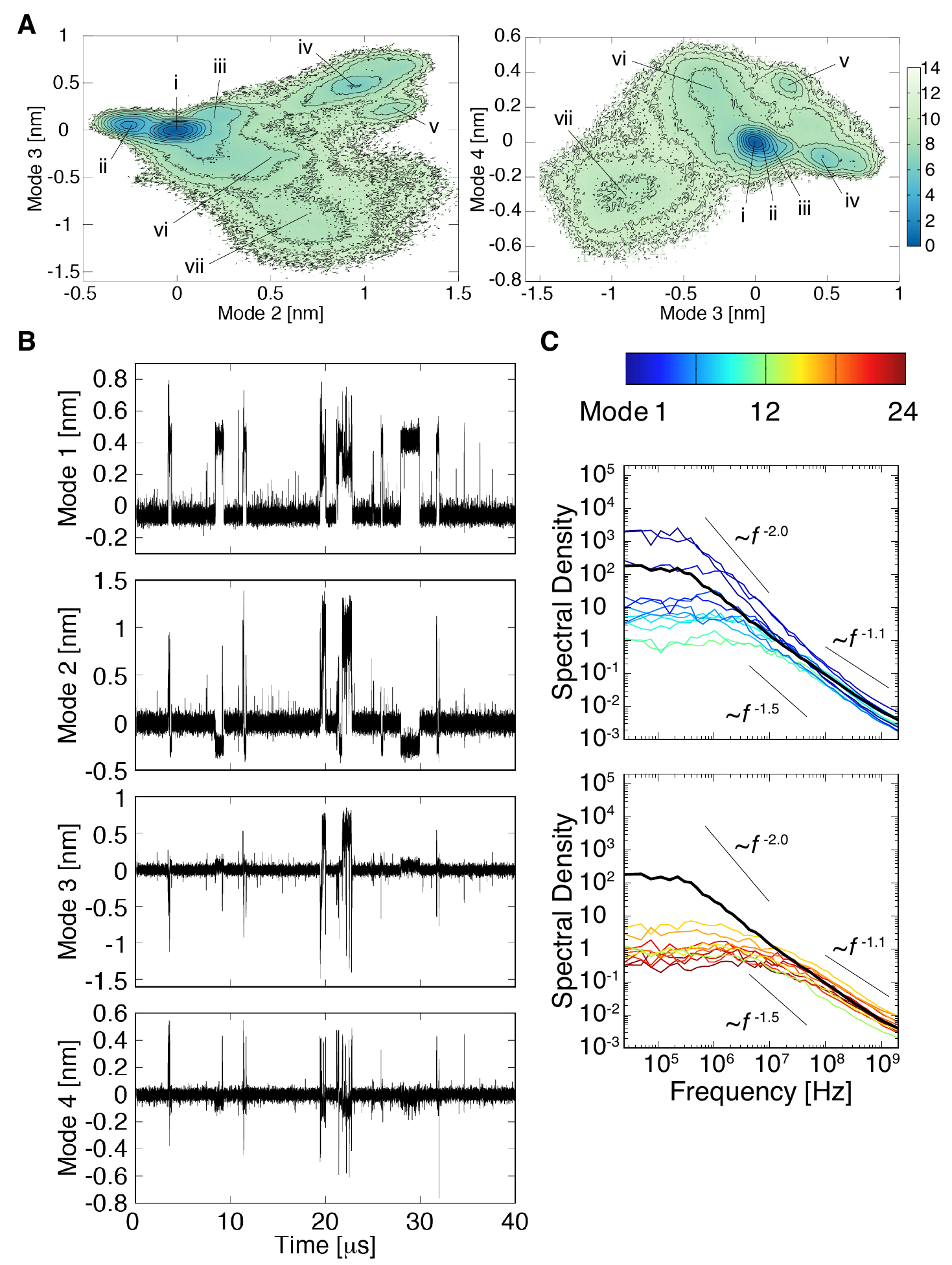}
\caption{Relaxation mode analysis (RMA) of Chignolin using the coordinates of C$\alpha$ atoms.
Parameters were set as $t_0 = 0.5$~ns and $\tau = 0.1$~ns.
(A)~Free energy map of  relaxation modes (RMs).
Snapshots of protein conformations corresponding to the states (i)-(vii) are shown in Fig.~2A (main text).
(B)~Time series of RMs.
The coordinates used for the analysis is the same as those in Fig.~1A (main text).
(B)~Ensemble averaged PSDs for 24 RMs.
Different colored lines represent the RM1 to RM24. 
The black bold line shows the cumulative PSD summed over 24 individual PSDs of each RM.}
\label{fig_FS_traj_rma_50_10}
\end{center}
\end{figure*}

\begin{figure*}[h]
\begin{center}
\includegraphics[width=130 mm,bb= 0 0 408 421]{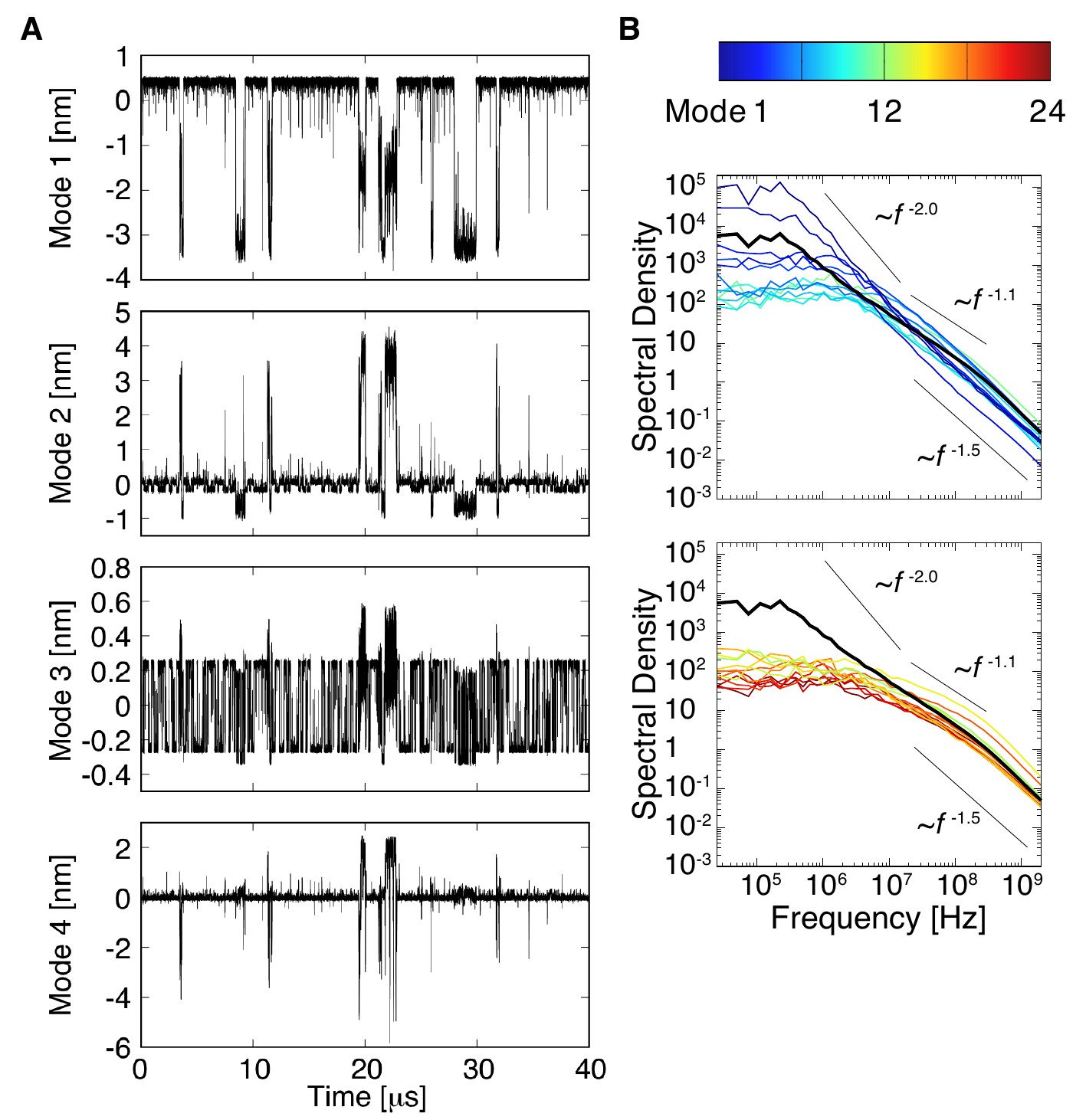}
\caption{RMA of Chignolin using the coordinates of heavy atoms.
Parameters were set as $t_0 = 0$~ns and $\tau = 0.1$~ns.
(A)~Time series of the RMs.
The coordinates used for the analysis is the same as those in Fig.~1A (main text).
RMA using the heavy atoms includes the dynamical modes of side chains. 
The time series of RM1 and RM2 are similar to those using the C$\alpha$ atoms.
The time series of RM3 corresponds to the rotational motion of the side chain of amino acid residues T2.
The time series of RM4 is similar to that of RM3 using the C$\alpha$ atoms.
(B)~Ensemble averaged PSDs for 24 RMs.
Different colored lines represent the RM1 to RM24. 
The black bold line shows the cumulative PSD summed over 24 individual PSDs of each RM.}
\label{fig_FS_traj_rma_h_0_10}
\end{center}
\end{figure*}

\begin{figure*}[h]
\begin{center}
\includegraphics[width=130 mm,bb= 0 0 408 416]{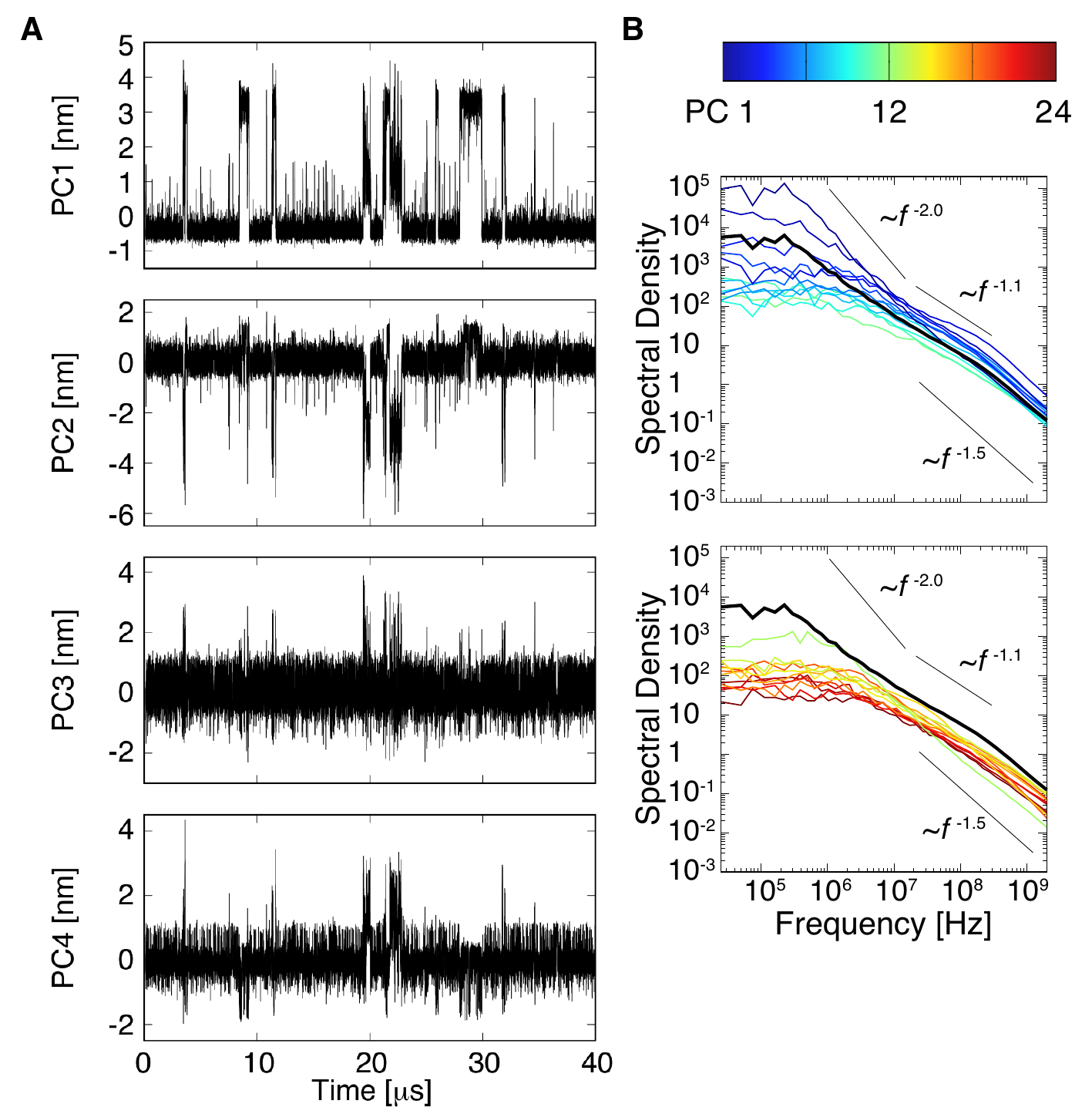}
\caption{Principal component analysis (PCA) of Chignolin using the coordinates of heavy atoms.
(A)~Time series of the principal components (PCs).
The coordinates used for the analysis is the same as those in Fig.~1A (main text).
(B)~Ensemble averaged PSDs for 24 PCs.
Different colored lines represent the PC1 to PC24. 
The black bold line shows the cumulative PSD summed over 24 individual PSDs of each PC.}
\label{fig_FS_traj_PCA_h}
\end{center}
\end{figure*}

\begin{figure*}[h]
\begin{center}
\includegraphics[width=150 mm,bb= 0 0 452 356]{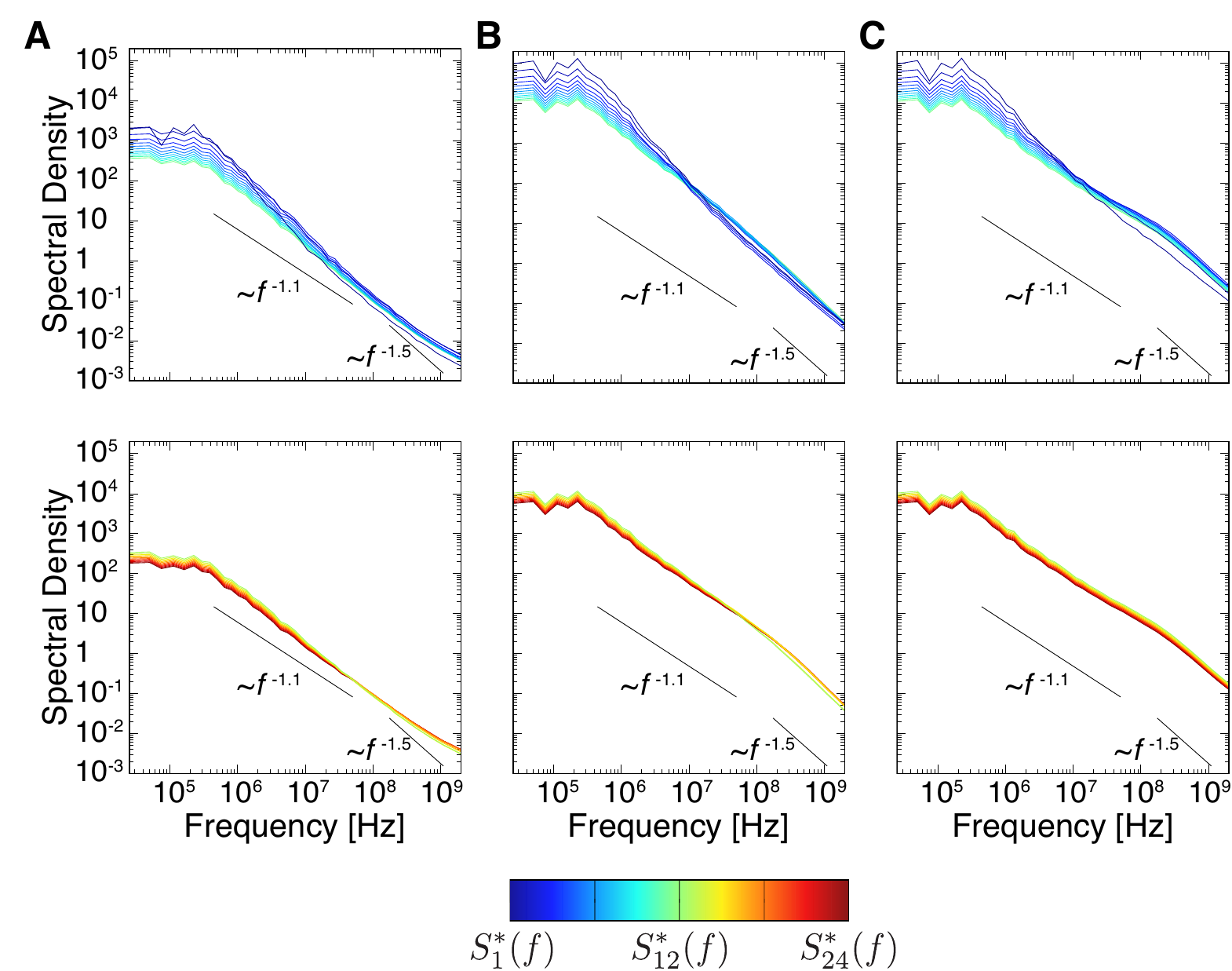}
\caption{Convergence of the cumulative PSDs summed over the PSDs of RM/PC.
(A)~RMA using Chignolin C$\alpha$ atoms ($t_0 = 0.5$~ns and $\tau = 0.1$~ns.), (B)~RMA using Chignolin heavy atoms ($t_0 = 0$~ns and $\tau = 0.1$~ns), and (C)~PCA using Chignolin heavy atoms.
Different colored lines represent the cumulative PSDs $S_n^*(f)$ summed over the PSDs of RM/PC, $S_n^*(f) = \sum^n_{i=1} S_i(f)$, where $S_i(f)$ is the PSD of $i$th RM/PC.
In the RMA using heavy atoms, transition of the power law exponents between $-1.1$ and $-1.5$ approximately appears after sum of 15 RMs.
In contrast, the PSDs of PCA show the transition after sum of a few PCs.
This means that RMA could decompose the modes of dynamics in more detail than PCA.}
\label{fig_FS_rma_PCA_fftw}
\end{center}
\end{figure*}

\begin{figure*}[h]
\begin{center}
\includegraphics[width=175 mm,bb= 0 0 754 221]{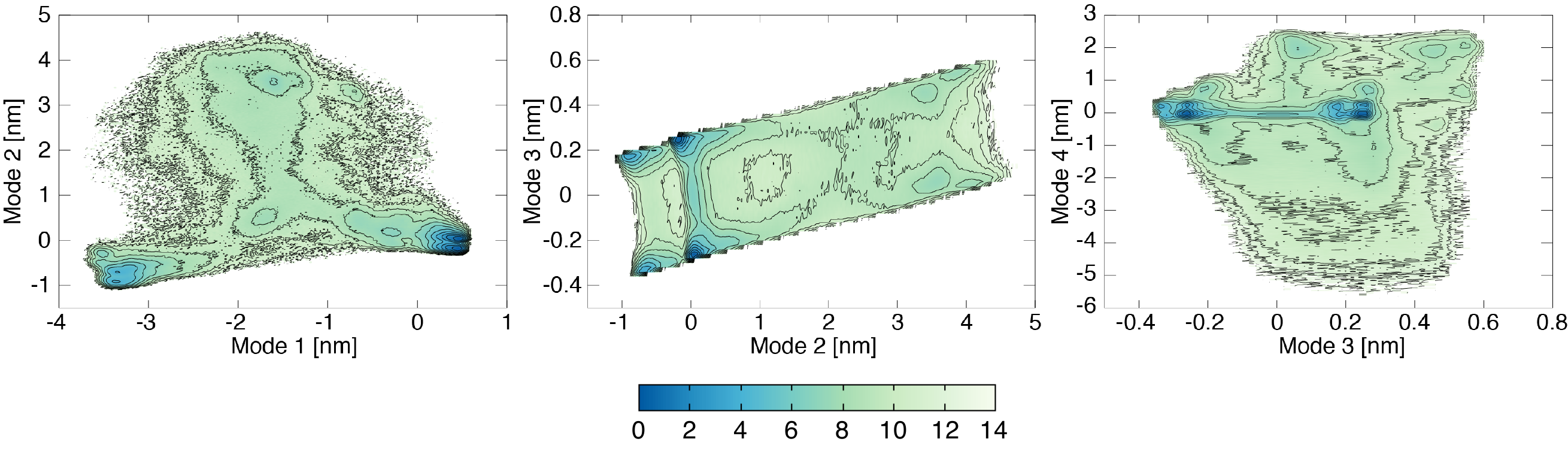}
\caption{Free energy maps of relaxation modes of Chignolin using the coordinates of heavy atoms.}
\label{fig_FS_rma_density_map_chignolin_h_0_10}
\end{center}
\end{figure*}

\begin{figure*}[h]
\begin{center}
\includegraphics[width=60 mm,bb= 0 0 162 150]{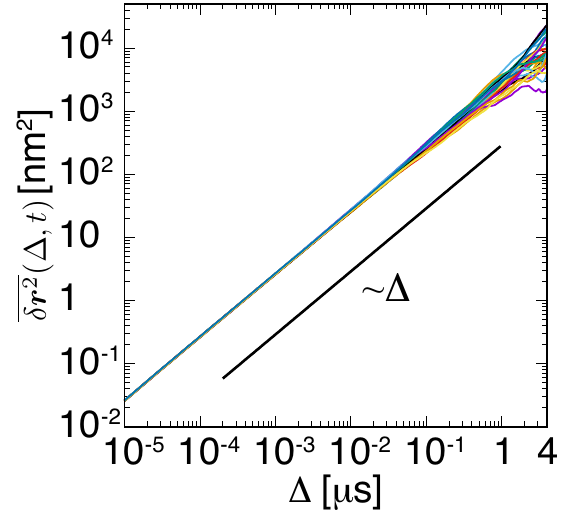}
\caption{Time-averaged mean squared displacements (TMSDs) of Chignolin.
30 trajectories divided from five runs of 40~$\mu$s simulation with a time window of $t = 6$~${\rm \mu}$s (measurement time) are shown.
The black solid line is shown for reference.}
\label{fig_diffusion}
\end{center}
\end{figure*}

\begin{figure*}[h]
\begin{center}
\includegraphics[width=85 mm,bb= 0 0 377 834]{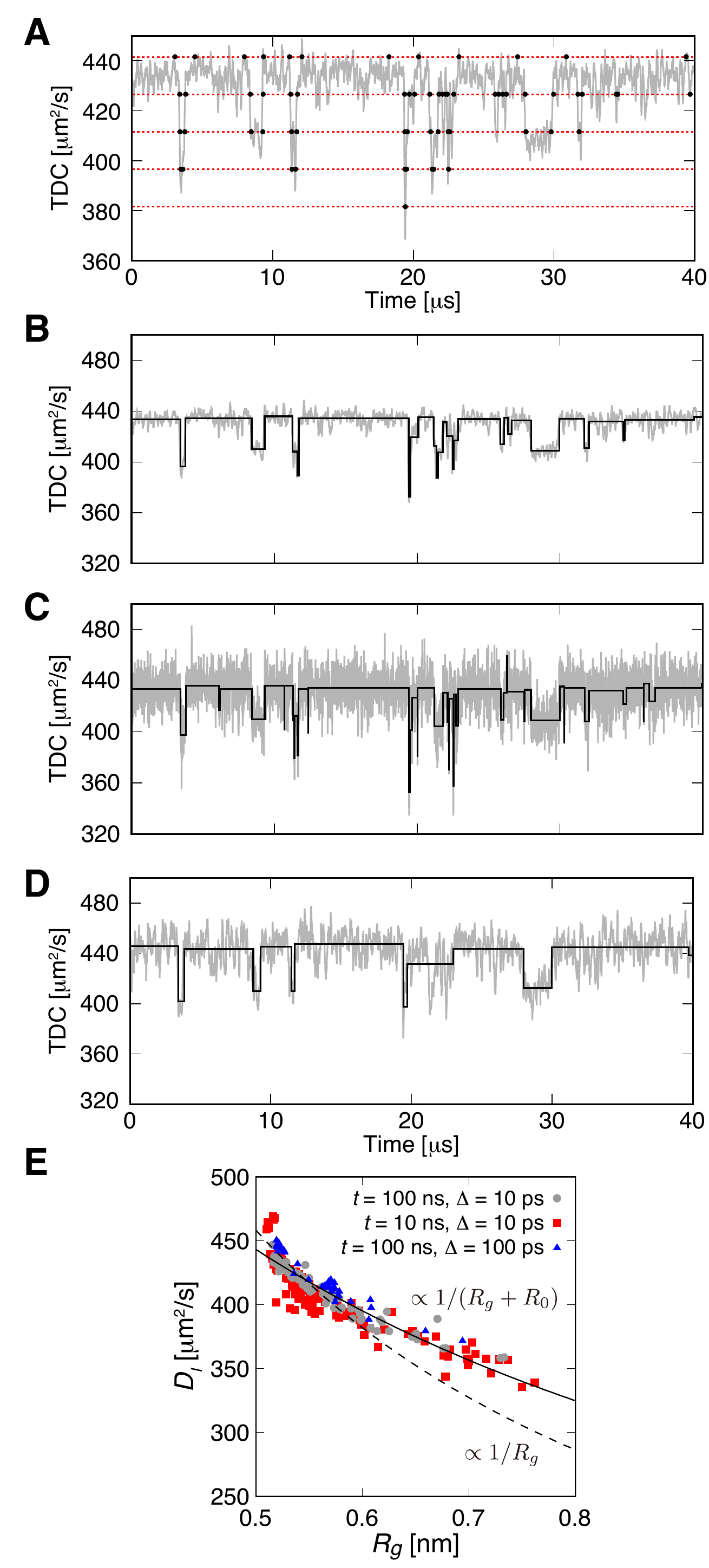}
\caption{Detection of the transition times of diffusivity of Chignolin.
(A)~Time series of temporal diffusion coefficient (TDC) with parameters $t=100$~ns, $\Delta=10$~ps.
The coordinates used for the analysis are the same as those in Fig.~1A (main text).
Thresholds $D_{\rm eff}^w$ are shown as red dashed lines.
Transition times for each $D_{\rm eff}^w$ after the correction with a statistical test are shown as circle symbols.
Time series of TDC (thin line) with the instantaneous diffusion coefficients $D_I$ (thick lines) with detected parameters: (B)~$t=100$~ns, $\Delta=10$~ps, (C)~$t=10$~ns, $\Delta=10$~ps, and (D)~$t=100$~ns, $\Delta=100$~ps.
(E)~Correlation between the mean $R_g$ and $D_I$ in each diffusive state.
Different colored symbols represent the correlation for different parameters.
The dashed and solid lines are shown as a reference of $D \propto 1/R_g$ and $D \propto 1/(R_g + R_0)$ with $R_0 = 0.3$ nm, respectively.
The correlation shows the same trend for these parameter sets.}
\label{fig_diffusion}
\end{center}
\end{figure*}

\begin{figure*}[h]
\begin{center}
\includegraphics[width=120 mm,bb= 0 0 1143 1098]{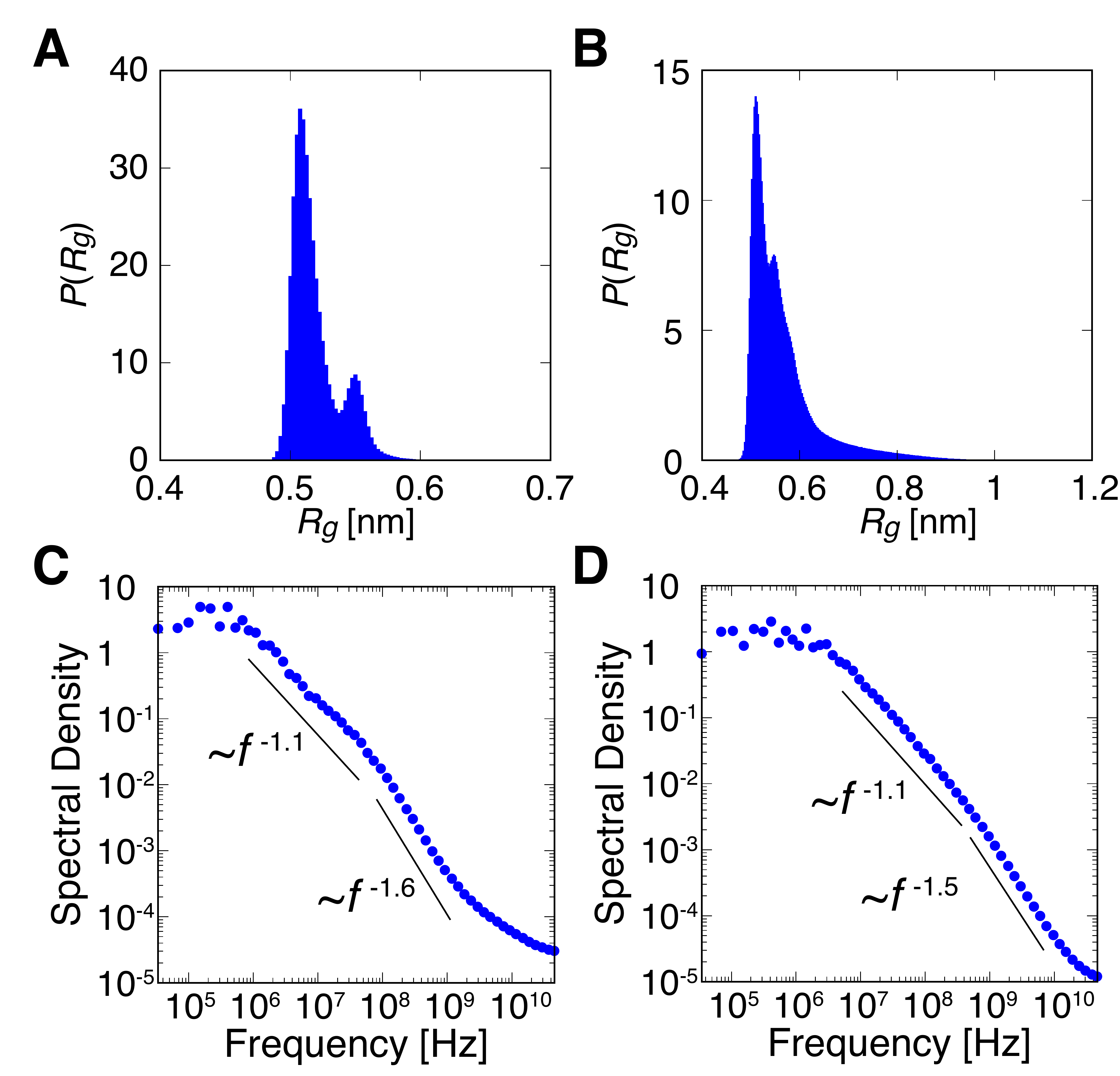}
\caption{Long-term correlation in the conformational fluctuation of Chignolin at different temperature and pressure conditions.
Probability density functions (PDFs) of $R_g$ (A)~at 280K and 0.1~MPa and (B)~at 400~K and 400~MPa.
Ensemble-averaged PSDs of $R_g$ (C)~at 280~K and 0.1~MPa and (D)~at 400~K and 400~MPa.
Solid lines are shown as a reference for power-law decays in higher and lower frequencies.
At the high temperature and pressure condition, the transition frequency }
\label{fig_FS_radius_gyration_fft}
\end{center}
\end{figure*}

\begin{figure*}[h]
\begin{center}
\includegraphics[width=175 mm,bb= 0 0 2248 1175]{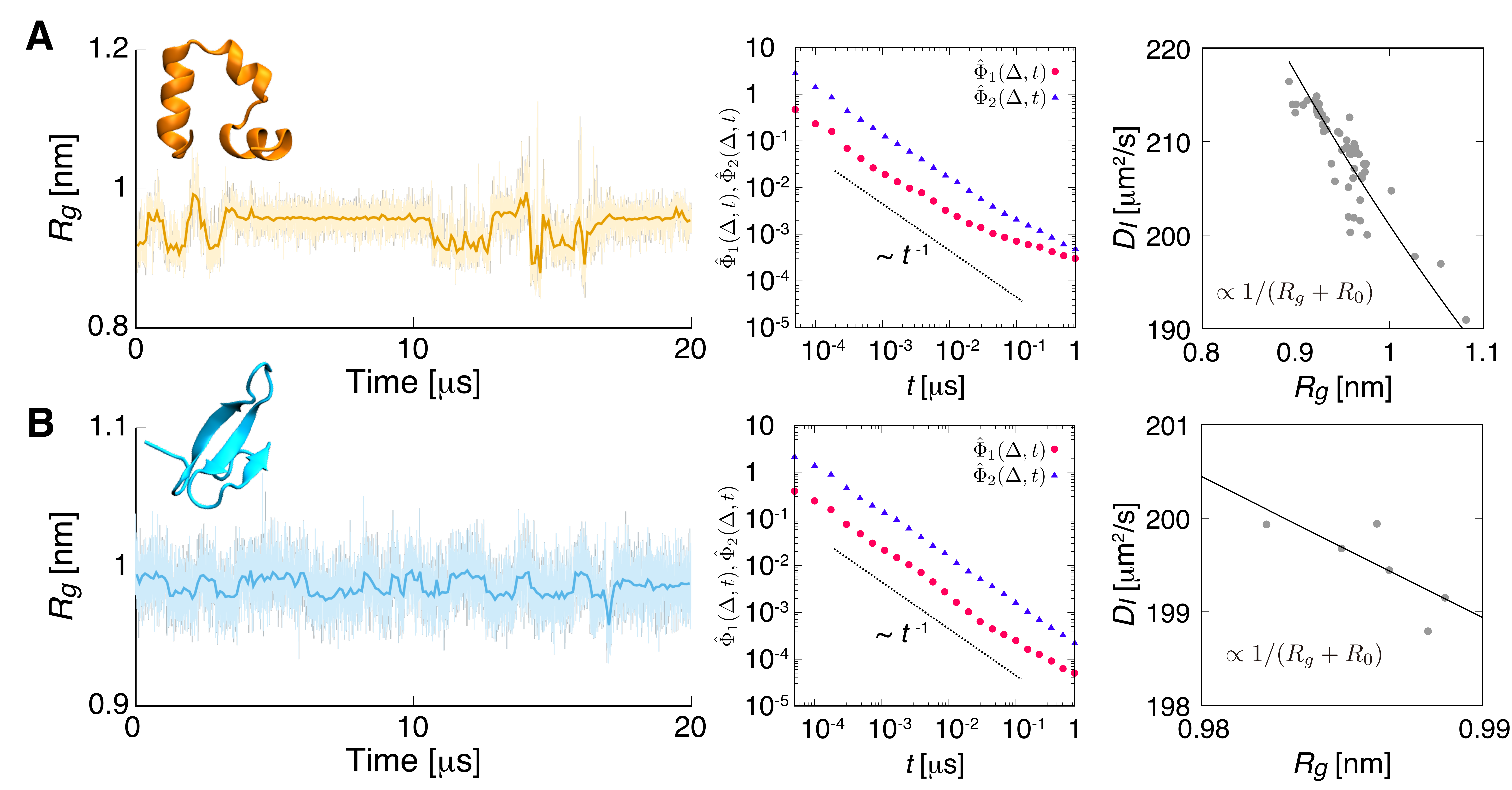}
\caption{Fluctuating diffusivity of (A)~Villin and (B)~WW domain of Pin1.
(Left)~Time series of the gyration radius $R_g$.
Thin and thick lines represent unsmoothed original values every 1~ns and smoothed moving average with 100~ns averaging window, respectively.
(Middle)~Normalized magnitude $\hat{\Phi}_1(\Delta, t)$ and orientation $\hat{\Phi}_2(\Delta,t)$ correlation functions.
Note that, in our simulations crossover of magnitude correlation was not observed for the longer timescale of the proteins.
(Right)~Correlation between the mean $R_g$ and the instantaneous diffusion coefficient $D_I$, where $t=100$~ns and $\Delta=10$~ps were used to obtain the TDCs.
The solid line is shown as a reference with $R_0 = 0.3$ nm.}
\label{fig_Figure5}
\end{center}
\end{figure*}

\begin{figure*}[h]
\begin{center}
\includegraphics[width=60 mm,bb= 0 0 177 155]{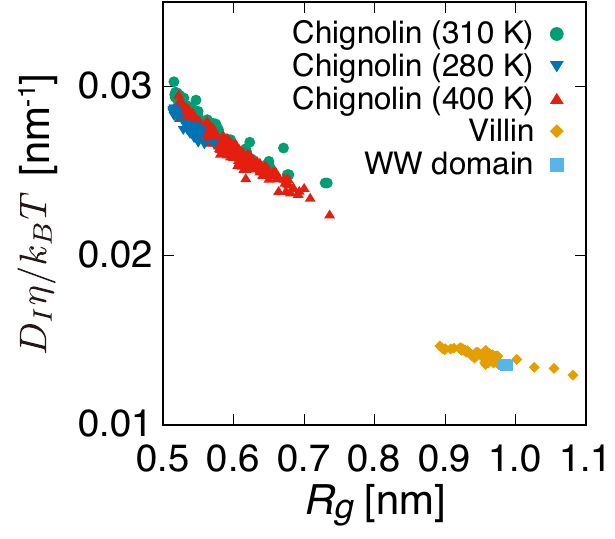}
\caption{Correlations between the mean $R_g$ and $D_I \eta / k_b T$ of Chignolin, Villin, and WW domain of Pin1 at $T=$310~K and 0.1~MPa, and Chignolin at 280K and 0.1~MPa and at 400~K and 400~MPa.
The values of the viscosity $\eta$ at different temperature and pressure conditions were calculated from the bulk TIP3P water systems.
$D_I \eta/k_BT$ for Chignolin at different temperature and pressure conditions are consistent, i.e. $D_I$ is proportional to $T/\eta$.}
\label{fig_Figure5}
\end{center}
\end{figure*}

\begin{figure*}[h]
\begin{center}
\includegraphics[width=60 mm,bb= 0 0 170 150]{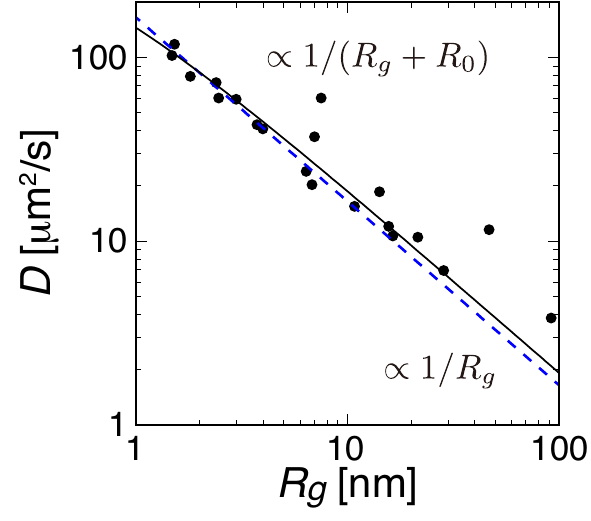}
\caption{Correlations between $R_g$ and diffusion coefficient $D$ for a variety of proteins at 293 K in dilute solution (data are obtained from Table 1 in Ref.~\cite{HeNiemeyer2003}).
The original experimental data are in Refs.~\cite{TynGusek1990, DurchschlagZipper1997}.
Different symbols represent the diffusion coefficient of different proteins.
The dashed and solid lines are shown as a reference of $D \propto 1/R_g$ and $D \propto 1/(R_g + R_0)$ with $R_0 = 0.3$ nm, respectively.
The SE-type relation $D \propto 1/(R_g + R_0)$ holds for the large variety of proteins, particularly for those with $R_g \gg R_0$.}
\label{fig_Figure5}
\end{center}
\end{figure*}


\begin{thebibliography}{62}%
\makeatletter
\providecommand \@ifxundefined [1]{%
 \@ifx{#1\undefined}
}%
\providecommand \@ifnum [1]{%
 \ifnum #1\expandafter \@firstoftwo
 \else \expandafter \@secondoftwo
 \fi
}%
\providecommand \@ifx [1]{%
 \ifx #1\expandafter \@firstoftwo
 \else \expandafter \@secondoftwo
 \fi
}%
\providecommand \natexlab [1]{#1}%
%
\providecommand \bibnamefont  [1]{#1}%
\providecommand \bibfnamefont [1]{#1}%
\providecommand \citenamefont [1]{#1}%
\providecommand \href@noop [0]{\@secondoftwo}%
\providecommand \href [0]{\begingroup \@sanitize@url \@href}%
\providecommand \@href[1]{\@@startlink{#1}\@@href}%
\providecommand \@@href[1]{\endgroup#1\@@endlink}%
\providecommand \@sanitize@url [0]{\catcode `\\12\catcode `\$12\catcode
  `\&12\catcode `\#12\catcode `\^12\catcode `\_12\catcode `\%12\relax}%
\providecommand \@@startlink[1]{}%
\providecommand \@@endlink[0]{}%
\providecommand \url  [0]{\begingroup\@sanitize@url \@url }%
\providecommand \@url [1]{\endgroup\@href {#1}{\urlprefix }}%
\providecommand \urlprefix  [0]{URL }%
%
\providecommand \doibase [0]{http://dx.doi.org/}%
\providecommand \selectlanguage [0]{\@gobble}%
\providecommand \bibinfo  [0]{\@secondoftwo}%
\providecommand \bibfield  [0]{\@secondoftwo}%
%
\providecommand \BibitemOpen [0]{}%
%
%
%
\providecommand \BibitemShut  [1]{\csname bibitem#1\endcsname}%
\let\auto@bib@innerbib\@empty
\bibitem [{\citenamefont {Young}\ \emph {et~al.}(1980)\citenamefont {Young},
  \citenamefont {Carroad},\ and\ \citenamefont {Bell}}]{YoungCarroadBell1980}%
  \BibitemOpen
  \bibfield  {author} {\bibinfo {author} {\bibfnamefont {M.~E.}\ \bibnamefont
  {Young}}, \bibinfo {author} {\bibfnamefont {P.~A.}\ \bibnamefont {Carroad}},
  \ and\ \bibinfo {author} {\bibfnamefont {R.~L.}\ \bibnamefont {Bell}},\
  }\href@noop {} {\bibfield  {journal} {\bibinfo  {journal} {Biotechnol.
  Bioeng.}\ }\textbf {\bibinfo {volume} {22}},\ \bibinfo {pages} {947}
  (\bibinfo {year} {1980})}\BibitemShut {NoStop}%
\bibitem [{\citenamefont {He}\ and\ \citenamefont
  {Niemeyer}(2003)}]{HeNiemeyer2003}%
  \BibitemOpen
  \bibfield  {author} {\bibinfo {author} {\bibfnamefont {L.}~\bibnamefont
  {He}}\ and\ \bibinfo {author} {\bibfnamefont {B.}~\bibnamefont {Niemeyer}},\
  }\href@noop {} {\bibfield  {journal} {\bibinfo  {journal} {Biotechnol.
  Prog.}\ }\textbf {\bibinfo {volume} {19}},\ \bibinfo {pages} {544} (\bibinfo
  {year} {2003})}\BibitemShut {NoStop}%
\bibitem [{\citenamefont {Tyn}\ and\ \citenamefont
  {Gusek}(1990)}]{TynGusek1990}%
  \BibitemOpen
  \bibfield  {author} {\bibinfo {author} {\bibfnamefont {M.~T.}\ \bibnamefont
  {Tyn}}\ and\ \bibinfo {author} {\bibfnamefont {T.~W.}\ \bibnamefont
  {Gusek}},\ }\href@noop {} {\bibfield  {journal} {\bibinfo  {journal}
  {Biotechnol. Bioeng.}\ }\textbf {\bibinfo {volume} {35}},\ \bibinfo {pages}
  {327} (\bibinfo {year} {1990})}\BibitemShut {NoStop}%
\bibitem [{\citenamefont {Halle}\ and\ \citenamefont
  {Davidovic}(2003)}]{HalleDavidovic2003}%
  \BibitemOpen
  \bibfield  {author} {\bibinfo {author} {\bibfnamefont {B.}~\bibnamefont
  {Halle}}\ and\ \bibinfo {author} {\bibfnamefont {M.}~\bibnamefont
  {Davidovic}},\ }\href@noop {} {\bibfield  {journal} {\bibinfo  {journal}
  {Proc. Natl. Acad. Sci. USA}\ }\textbf {\bibinfo {volume} {100}},\ \bibinfo
  {pages} {12135} (\bibinfo {year} {2003})}\BibitemShut {NoStop}%
\bibitem [{\citenamefont {Minton}(2001)}]{Minton2001}%
  \BibitemOpen
  \bibfield  {author} {\bibinfo {author} {\bibfnamefont {A.~P.}\ \bibnamefont
  {Minton}},\ }\href@noop {} {\bibfield  {journal} {\bibinfo  {journal} {J.
  Biol. Chem.}\ }\textbf {\bibinfo {volume} {276}},\ \bibinfo {pages} {10577}
  (\bibinfo {year} {2001})}\BibitemShut {NoStop}%
\bibitem [{\citenamefont {Metzler}\ \emph {et~al.}(2016)\citenamefont
  {Metzler}, \citenamefont {Jeon},\ and\ \citenamefont
  {Cherstvy}}]{MetzlerJeonCherstvy2016}%
  \BibitemOpen
  \bibfield  {author} {\bibinfo {author} {\bibfnamefont {R.}~\bibnamefont
  {Metzler}}, \bibinfo {author} {\bibfnamefont {J.-H.}\ \bibnamefont {Jeon}}, \
  and\ \bibinfo {author} {\bibfnamefont {A.~G.}\ \bibnamefont {Cherstvy}},\
  }\href@noop {} {\bibfield  {journal} {\bibinfo  {journal} {Biochim. Biophys.
  Acta}\ }\textbf {\bibinfo {volume} {1858}},\ \bibinfo {pages} {2451}
  (\bibinfo {year} {2016})}\BibitemShut {NoStop}%
\bibitem [{\citenamefont {Javanainen}\ \emph {et~al.}(2017)\citenamefont
  {Javanainen}, \citenamefont {Martinez-Seara}, \citenamefont {Metzler},\ and\
  \citenamefont
  {Vattulainen}}]{JavanainenMartinez-SearaMetzlerVattulainen2017}%
  \BibitemOpen
  \bibfield  {author} {\bibinfo {author} {\bibfnamefont {M.}~\bibnamefont
  {Javanainen}}, \bibinfo {author} {\bibfnamefont {H.}~\bibnamefont
  {Martinez-Seara}}, \bibinfo {author} {\bibfnamefont {R.}~\bibnamefont
  {Metzler}}, \ and\ \bibinfo {author} {\bibfnamefont {I.}~\bibnamefont
  {Vattulainen}},\ }\href@noop {} {\bibfield  {journal} {\bibinfo  {journal}
  {J. Phys. Chem. Lett.}\ }\textbf {\bibinfo {volume} {8}},\ \bibinfo {pages}
  {4308} (\bibinfo {year} {2017})}\BibitemShut {NoStop}%
\bibitem [{\citenamefont {Wei{\ss}}\ \emph {et~al.}(2013)\citenamefont
  {Wei{\ss}}, \citenamefont {Neef}, \citenamefont {Van}, \citenamefont
  {Kramer}, \citenamefont {Gregor},\ and\ \citenamefont
  {Enderlein}}]{WeisNeefVanKramerGregorEnderlein2013}%
  \BibitemOpen
  \bibfield  {author} {\bibinfo {author} {\bibfnamefont {K.}~\bibnamefont
  {Wei{\ss}}}, \bibinfo {author} {\bibfnamefont {A.}~\bibnamefont {Neef}},
  \bibinfo {author} {\bibfnamefont {Q.}~\bibnamefont {Van}}, \bibinfo {author}
  {\bibfnamefont {S.}~\bibnamefont {Kramer}}, \bibinfo {author} {\bibfnamefont
  {I.}~\bibnamefont {Gregor}}, \ and\ \bibinfo {author} {\bibfnamefont
  {J.}~\bibnamefont {Enderlein}},\ }\href@noop {} {\bibfield  {journal}
  {\bibinfo  {journal} {Biophys. J.}\ }\textbf {\bibinfo {volume} {105}},\
  \bibinfo {pages} {455} (\bibinfo {year} {2013})}\BibitemShut {NoStop}%
\bibitem [{\citenamefont {Yamamoto}\ and\ \citenamefont
  {Onuki}(1998)}]{YamamotoOnuki1998}%
  \BibitemOpen
  \bibfield  {author} {\bibinfo {author} {\bibfnamefont {R.}~\bibnamefont
  {Yamamoto}}\ and\ \bibinfo {author} {\bibfnamefont {A.}~\bibnamefont
  {Onuki}},\ }\href {\doibase 10.1103/PhysRevLett.81.4915} {\bibfield
  {journal} {\bibinfo  {journal} {Phys. Rev. Lett.}\ }\textbf {\bibinfo
  {volume} {81}},\ \bibinfo {pages} {4915} (\bibinfo {year}
  {1998})}\BibitemShut {NoStop}%
\bibitem [{\citenamefont {Wang}\ \emph {et~al.}(2009)\citenamefont {Wang},
  \citenamefont {Anthony}, \citenamefont {Bae},\ and\ \citenamefont
  {Granick}}]{WangAnthonyBaeGranick2009}%
  \BibitemOpen
  \bibfield  {author} {\bibinfo {author} {\bibfnamefont {B.}~\bibnamefont
  {Wang}}, \bibinfo {author} {\bibfnamefont {S.~M.}\ \bibnamefont {Anthony}},
  \bibinfo {author} {\bibfnamefont {S.~C.}\ \bibnamefont {Bae}}, \ and\
  \bibinfo {author} {\bibfnamefont {S.}~\bibnamefont {Granick}},\ }\href@noop
  {} {\bibfield  {journal} {\bibinfo  {journal} {Proc. Natl. Acad. Sci. USA}\
  }\textbf {\bibinfo {volume} {106}},\ \bibinfo {pages} {15160} (\bibinfo
  {year} {2009})}\BibitemShut {NoStop}%
\bibitem [{\citenamefont {Wang}\ \emph {et~al.}(2012)\citenamefont {Wang},
  \citenamefont {Kuo}, \citenamefont {Bae},\ and\ \citenamefont
  {Granick}}]{WangKuoBaeGranick2012}%
  \BibitemOpen
  \bibfield  {author} {\bibinfo {author} {\bibfnamefont {B.}~\bibnamefont
  {Wang}}, \bibinfo {author} {\bibfnamefont {J.}~\bibnamefont {Kuo}}, \bibinfo
  {author} {\bibfnamefont {S.~C.}\ \bibnamefont {Bae}}, \ and\ \bibinfo
  {author} {\bibfnamefont {S.}~\bibnamefont {Granick}},\ }\href@noop {}
  {\bibfield  {journal} {\bibinfo  {journal} {Nat. Mater.}\ }\textbf {\bibinfo
  {volume} {11}},\ \bibinfo {pages} {481} (\bibinfo {year} {2012})}\BibitemShut
  {NoStop}%
\bibitem [{\citenamefont {Serg{\'e}}\ \emph {et~al.}(2008)\citenamefont
  {Serg{\'e}}, \citenamefont {Bertaux}, \citenamefont {Rigneault},\ and\
  \citenamefont {Marguet}}]{SergeBertauxRigneaultMarguet2008}%
  \BibitemOpen
  \bibfield  {author} {\bibinfo {author} {\bibfnamefont {A.}~\bibnamefont
  {Serg{\'e}}}, \bibinfo {author} {\bibfnamefont {N.}~\bibnamefont {Bertaux}},
  \bibinfo {author} {\bibfnamefont {H.}~\bibnamefont {Rigneault}}, \ and\
  \bibinfo {author} {\bibfnamefont {D.}~\bibnamefont {Marguet}},\ }\href@noop
  {} {\bibfield  {journal} {\bibinfo  {journal} {Nat. Methods}\ }\textbf
  {\bibinfo {volume} {5}},\ \bibinfo {pages} {687} (\bibinfo {year}
  {2008})}\BibitemShut {NoStop}%
\bibitem [{\citenamefont {Manzo}\ \emph {et~al.}(2015)\citenamefont {Manzo},
  \citenamefont {Torreno-Pina}, \citenamefont {Massignan}, \citenamefont
  {Lapeyre}, \citenamefont {Lewenstein},\ and\ \citenamefont
  {Garcia~Parajo}}]{ManzoTorreno-PinaMassignanLapeyreLewensteinGarcia2015}%
  \BibitemOpen
  \bibfield  {author} {\bibinfo {author} {\bibfnamefont {C.}~\bibnamefont
  {Manzo}}, \bibinfo {author} {\bibfnamefont {J.~A.}\ \bibnamefont
  {Torreno-Pina}}, \bibinfo {author} {\bibfnamefont {P.}~\bibnamefont
  {Massignan}}, \bibinfo {author} {\bibfnamefont {G.~J.}\ \bibnamefont
  {Lapeyre}}, \bibinfo {author} {\bibfnamefont {M.}~\bibnamefont {Lewenstein}},
  \ and\ \bibinfo {author} {\bibfnamefont {M.~F.}\ \bibnamefont
  {Garcia~Parajo}},\ }\href {\doibase 10.1103/PhysRevX.5.011021} {\bibfield
  {journal} {\bibinfo  {journal} {Phys. Rev. X}\ }\textbf {\bibinfo {volume}
  {5}},\ \bibinfo {pages} {011021} (\bibinfo {year} {2015})}\BibitemShut
  {NoStop}%
\bibitem [{\citenamefont {Yamamoto}\ \emph {et~al.}(2015)\citenamefont
  {Yamamoto}, \citenamefont {Kalli}, \citenamefont {Akimoto}, \citenamefont
  {Yasuoka},\ and\ \citenamefont
  {Sansom}}]{YamamotoKalliAkimotoYasuokaSansom2015}%
  \BibitemOpen
  \bibfield  {author} {\bibinfo {author} {\bibfnamefont {E.}~\bibnamefont
  {Yamamoto}}, \bibinfo {author} {\bibfnamefont {A.~C.}\ \bibnamefont {Kalli}},
  \bibinfo {author} {\bibfnamefont {T.}~\bibnamefont {Akimoto}}, \bibinfo
  {author} {\bibfnamefont {K.}~\bibnamefont {Yasuoka}}, \ and\ \bibinfo
  {author} {\bibfnamefont {M.~S.~P.}\ \bibnamefont {Sansom}},\ }\href@noop {}
  {\bibfield  {journal} {\bibinfo  {journal} {Sci. Rep.}\ }\textbf {\bibinfo
  {volume} {5}},\ \bibinfo {pages} {18245} (\bibinfo {year}
  {2015})}\BibitemShut {NoStop}%
\bibitem [{\citenamefont {Jeon}\ \emph {et~al.}(2016)\citenamefont {Jeon},
  \citenamefont {Javanainen}, \citenamefont {Martinez-Seara}, \citenamefont
  {Metzler},\ and\ \citenamefont
  {Vattulainen}}]{JeonJavanainenMartinez-SearaMetzlerVattulainen2016}%
  \BibitemOpen
  \bibfield  {author} {\bibinfo {author} {\bibfnamefont {J.-H.}\ \bibnamefont
  {Jeon}}, \bibinfo {author} {\bibfnamefont {M.}~\bibnamefont {Javanainen}},
  \bibinfo {author} {\bibfnamefont {H.}~\bibnamefont {Martinez-Seara}},
  \bibinfo {author} {\bibfnamefont {R.}~\bibnamefont {Metzler}}, \ and\
  \bibinfo {author} {\bibfnamefont {I.}~\bibnamefont {Vattulainen}},\ }\href
  {\doibase 10.1103/PhysRevX.6.021006} {\bibfield  {journal} {\bibinfo
  {journal} {Phys. Rev. X}\ }\textbf {\bibinfo {volume} {6}},\ \bibinfo {pages}
  {021006} (\bibinfo {year} {2016})}\BibitemShut {NoStop}%
\bibitem [{\citenamefont {He}\ \emph {et~al.}(2016)\citenamefont {He},
  \citenamefont {Song}, \citenamefont {Su}, \citenamefont {Geng}, \citenamefont
  {Ackerson}, \citenamefont {Peng},\ and\ \citenamefont
  {Tong}}]{HeSongSuGengAckersonPengTong2016}%
  \BibitemOpen
  \bibfield  {author} {\bibinfo {author} {\bibfnamefont {W.}~\bibnamefont
  {He}}, \bibinfo {author} {\bibfnamefont {H.}~\bibnamefont {Song}}, \bibinfo
  {author} {\bibfnamefont {Y.}~\bibnamefont {Su}}, \bibinfo {author}
  {\bibfnamefont {L.}~\bibnamefont {Geng}}, \bibinfo {author} {\bibfnamefont
  {B.~J.}\ \bibnamefont {Ackerson}}, \bibinfo {author} {\bibfnamefont {H.~B.}\
  \bibnamefont {Peng}}, \ and\ \bibinfo {author} {\bibfnamefont
  {P.}~\bibnamefont {Tong}},\ }\href@noop {} {\bibfield  {journal} {\bibinfo
  {journal} {Nat. Commun.}\ }\textbf {\bibinfo {volume} {7}},\ \bibinfo {pages}
  {11701} (\bibinfo {year} {2016})}\BibitemShut {NoStop}%
\bibitem [{\citenamefont {Weron}\ \emph {et~al.}(2017)\citenamefont {Weron},
  \citenamefont {Burnecki}, \citenamefont {Akin}, \citenamefont {Sol{\'e}},
  \citenamefont {Balcerek}, \citenamefont {Tamkun},\ and\ \citenamefont
  {Krapf}}]{WeronBurneckiAkinSoleBalcerekTamkunKrapf2017}%
  \BibitemOpen
  \bibfield  {author} {\bibinfo {author} {\bibfnamefont {A.}~\bibnamefont
  {Weron}}, \bibinfo {author} {\bibfnamefont {K.}~\bibnamefont {Burnecki}},
  \bibinfo {author} {\bibfnamefont {E.~J.}\ \bibnamefont {Akin}}, \bibinfo
  {author} {\bibfnamefont {L.}~\bibnamefont {Sol{\'e}}}, \bibinfo {author}
  {\bibfnamefont {M.}~\bibnamefont {Balcerek}}, \bibinfo {author}
  {\bibfnamefont {M.~M.}\ \bibnamefont {Tamkun}}, \ and\ \bibinfo {author}
  {\bibfnamefont {D.}~\bibnamefont {Krapf}},\ }\href@noop {} {\bibfield
  {journal} {\bibinfo  {journal} {Sci. Rep.}\ }\textbf {\bibinfo {volume}
  {7}},\ \bibinfo {pages} {5404} (\bibinfo {year} {2017})}\BibitemShut
  {NoStop}%
\bibitem [{\citenamefont {Yamamoto}\ \emph {et~al.}(2017)\citenamefont
  {Yamamoto}, \citenamefont {Akimoto}, \citenamefont {Kalli}, \citenamefont
  {Yasuoka},\ and\ \citenamefont
  {Sansom}}]{YamamotoAkimotoKalliYasuokaSansom2017}%
  \BibitemOpen
  \bibfield  {author} {\bibinfo {author} {\bibfnamefont {E.}~\bibnamefont
  {Yamamoto}}, \bibinfo {author} {\bibfnamefont {T.}~\bibnamefont {Akimoto}},
  \bibinfo {author} {\bibfnamefont {A.~C.}\ \bibnamefont {Kalli}}, \bibinfo
  {author} {\bibfnamefont {K.}~\bibnamefont {Yasuoka}}, \ and\ \bibinfo
  {author} {\bibfnamefont {M.~S.~P.}\ \bibnamefont {Sansom}},\ }\href@noop {}
  {\bibfield  {journal} {\bibinfo  {journal} {Science Adv.}\ }\textbf {\bibinfo
  {volume} {3}},\ \bibinfo {pages} {e1601871} (\bibinfo {year}
  {2017})}\BibitemShut {NoStop}%
\bibitem [{\citenamefont {Lampo}\ \emph {et~al.}(2017)\citenamefont {Lampo},
  \citenamefont {Stylianidou}, \citenamefont {Backlund}, \citenamefont
  {Wiggins},\ and\ \citenamefont
  {Spakowitz}}]{LampoStylianidouBacklundWigginsSpakowitz2017}%
  \BibitemOpen
  \bibfield  {author} {\bibinfo {author} {\bibfnamefont {T.~J.}\ \bibnamefont
  {Lampo}}, \bibinfo {author} {\bibfnamefont {S.}~\bibnamefont {Stylianidou}},
  \bibinfo {author} {\bibfnamefont {M.~P.}\ \bibnamefont {Backlund}}, \bibinfo
  {author} {\bibfnamefont {P.~A.}\ \bibnamefont {Wiggins}}, \ and\ \bibinfo
  {author} {\bibfnamefont {A.~J.}\ \bibnamefont {Spakowitz}},\ }\href@noop {}
  {\bibfield  {journal} {\bibinfo  {journal} {Biophys. J.}\ }\textbf {\bibinfo
  {volume} {112}},\ \bibinfo {pages} {532} (\bibinfo {year}
  {2017})}\BibitemShut {NoStop}%
\bibitem [{\citenamefont {Cherstvy}\ \emph {et~al.}(2018)\citenamefont
  {Cherstvy}, \citenamefont {Nagel}, \citenamefont {Beta},\ and\ \citenamefont
  {Metzler}}]{CherstvyNagelBetaMetzler2018}%
  \BibitemOpen
  \bibfield  {author} {\bibinfo {author} {\bibfnamefont {A.~G.}\ \bibnamefont
  {Cherstvy}}, \bibinfo {author} {\bibfnamefont {O.}~\bibnamefont {Nagel}},
  \bibinfo {author} {\bibfnamefont {C.}~\bibnamefont {Beta}}, \ and\ \bibinfo
  {author} {\bibfnamefont {R.}~\bibnamefont {Metzler}},\ }\href@noop {}
  {\bibfield  {journal} {\bibinfo  {journal} {Phys. Chem. Chem. Phys.}\
  }\textbf {\bibinfo {volume} {20}},\ \bibinfo {pages} {23034} (\bibinfo {year}
  {2018})}\BibitemShut {NoStop}%
\bibitem [{\citenamefont {Massignan}\ \emph {et~al.}(2014)\citenamefont
  {Massignan}, \citenamefont {Manzo}, \citenamefont {Torreno-Pina},
  \citenamefont {Garc\'{\i}a-Parajo}, \citenamefont {Lewenstein},\ and\
  \citenamefont
  {Lapeyre}}]{MassignanManzoTorreno-PinaGarcia-ParajoLewensteinLapeyre2014}%
  \BibitemOpen
  \bibfield  {author} {\bibinfo {author} {\bibfnamefont {P.}~\bibnamefont
  {Massignan}}, \bibinfo {author} {\bibfnamefont {C.}~\bibnamefont {Manzo}},
  \bibinfo {author} {\bibfnamefont {J.~A.}\ \bibnamefont {Torreno-Pina}},
  \bibinfo {author} {\bibfnamefont {M.~F.}\ \bibnamefont {Garc\'{\i}a-Parajo}},
  \bibinfo {author} {\bibfnamefont {M.}~\bibnamefont {Lewenstein}}, \ and\
  \bibinfo {author} {\bibfnamefont {G.~J.}\ \bibnamefont {Lapeyre}},\ }\href
  {\doibase 10.1103/PhysRevLett.112.150603} {\bibfield  {journal} {\bibinfo
  {journal} {Phys. Rev. Lett.}\ }\textbf {\bibinfo {volume} {112}},\ \bibinfo
  {pages} {150603} (\bibinfo {year} {2014})}\BibitemShut {NoStop}%
\bibitem [{\citenamefont {Chubynsky}\ and\ \citenamefont
  {Slater}(2014)}]{ChubynskySlater2014}%
  \BibitemOpen
  \bibfield  {author} {\bibinfo {author} {\bibfnamefont {M.~V.}\ \bibnamefont
  {Chubynsky}}\ and\ \bibinfo {author} {\bibfnamefont {G.~W.}\ \bibnamefont
  {Slater}},\ }\href {\doibase 10.1103/PhysRevLett.113.098302} {\bibfield
  {journal} {\bibinfo  {journal} {Phys. Rev. Lett.}\ }\textbf {\bibinfo
  {volume} {113}},\ \bibinfo {pages} {098302} (\bibinfo {year}
  {2014})}\BibitemShut {NoStop}%
\bibitem [{\citenamefont {Uneyama}\ \emph {et~al.}(2015)\citenamefont
  {Uneyama}, \citenamefont {Miyaguchi},\ and\ \citenamefont
  {Akimoto}}]{UneyamaMiyaguchiAkimoto2015}%
  \BibitemOpen
  \bibfield  {author} {\bibinfo {author} {\bibfnamefont {T.}~\bibnamefont
  {Uneyama}}, \bibinfo {author} {\bibfnamefont {T.}~\bibnamefont {Miyaguchi}},
  \ and\ \bibinfo {author} {\bibfnamefont {T.}~\bibnamefont {Akimoto}},\ }\href
  {\doibase 10.1103/PhysRevE.92.032140} {\bibfield  {journal} {\bibinfo
  {journal} {Phys. Rev. E}\ }\textbf {\bibinfo {volume} {92}},\ \bibinfo
  {pages} {032140} (\bibinfo {year} {2015})}\BibitemShut {NoStop}%
\bibitem [{\citenamefont {Akimoto}\ and\ \citenamefont
  {Yamamoto}(2016{\natexlab{a}})}]{AkimotoYamamoto2016}%
  \BibitemOpen
  \bibfield  {author} {\bibinfo {author} {\bibfnamefont {T.}~\bibnamefont
  {Akimoto}}\ and\ \bibinfo {author} {\bibfnamefont {E.}~\bibnamefont
  {Yamamoto}},\ }\href {\doibase 10.1103/PhysRevE.93.062109} {\bibfield
  {journal} {\bibinfo  {journal} {Phys. Rev. E}\ }\textbf {\bibinfo {volume}
  {93}},\ \bibinfo {pages} {062109} (\bibinfo {year}
  {2016}{\natexlab{a}})}\BibitemShut {NoStop}%
\bibitem [{\citenamefont {Miyaguchi}\ \emph {et~al.}(2016)\citenamefont
  {Miyaguchi}, \citenamefont {Akimoto},\ and\ \citenamefont
  {Yamamoto}}]{MiyaguchiAkimotoYamamoto2016}%
  \BibitemOpen
  \bibfield  {author} {\bibinfo {author} {\bibfnamefont {T.}~\bibnamefont
  {Miyaguchi}}, \bibinfo {author} {\bibfnamefont {T.}~\bibnamefont {Akimoto}},
  \ and\ \bibinfo {author} {\bibfnamefont {E.}~\bibnamefont {Yamamoto}},\
  }\href {\doibase 10.1103/PhysRevE.94.012109} {\bibfield  {journal} {\bibinfo
  {journal} {Phys. Rev. E}\ }\textbf {\bibinfo {volume} {94}},\ \bibinfo
  {pages} {012109} (\bibinfo {year} {2016})}\BibitemShut {NoStop}%
\bibitem [{\citenamefont {Cherstvy}\ and\ \citenamefont
  {Metzler}(2016)}]{CherstvyMetzler2016}%
  \BibitemOpen
  \bibfield  {author} {\bibinfo {author} {\bibfnamefont {A.~G.}\ \bibnamefont
  {Cherstvy}}\ and\ \bibinfo {author} {\bibfnamefont {R.}~\bibnamefont
  {Metzler}},\ }\href@noop {} {\bibfield  {journal} {\bibinfo  {journal} {Phys.
  Chem. Chem. Phys.}\ }\textbf {\bibinfo {volume} {18}},\ \bibinfo {pages}
  {23840} (\bibinfo {year} {2016})}\BibitemShut {NoStop}%
\bibitem [{\citenamefont {Chechkin}\ \emph {et~al.}(2017)\citenamefont
  {Chechkin}, \citenamefont {Seno}, \citenamefont {Metzler},\ and\
  \citenamefont {Sokolov}}]{ChechkinSenoMetzlerSokolov2017}%
  \BibitemOpen
  \bibfield  {author} {\bibinfo {author} {\bibfnamefont {A.~V.}\ \bibnamefont
  {Chechkin}}, \bibinfo {author} {\bibfnamefont {F.}~\bibnamefont {Seno}},
  \bibinfo {author} {\bibfnamefont {R.}~\bibnamefont {Metzler}}, \ and\
  \bibinfo {author} {\bibfnamefont {I.~M.}\ \bibnamefont {Sokolov}},\ }\href
  {\doibase 10.1103/PhysRevX.7.021002} {\bibfield  {journal} {\bibinfo
  {journal} {Phys. Rev. X}\ }\textbf {\bibinfo {volume} {7}},\ \bibinfo {pages}
  {021002} (\bibinfo {year} {2017})}\BibitemShut {NoStop}%
\bibitem [{\citenamefont {Tyagi}\ and\ \citenamefont
  {Cherayil}(2017)}]{TyagiCherayil2017}%
  \BibitemOpen
  \bibfield  {author} {\bibinfo {author} {\bibfnamefont {N.}~\bibnamefont
  {Tyagi}}\ and\ \bibinfo {author} {\bibfnamefont {B.~J.}\ \bibnamefont
  {Cherayil}},\ }\href@noop {} {\bibfield  {journal} {\bibinfo  {journal} {J.
  Phys. Chem. B}\ }\textbf {\bibinfo {volume} {121}},\ \bibinfo {pages} {7204}
  (\bibinfo {year} {2017})}\BibitemShut {NoStop}%
\bibitem [{\citenamefont {Jain}\ and\ \citenamefont
  {Sebastian}(2018)}]{JainSebastian2018}%
  \BibitemOpen
  \bibfield  {author} {\bibinfo {author} {\bibfnamefont {R.}~\bibnamefont
  {Jain}}\ and\ \bibinfo {author} {\bibfnamefont {K.~L.}\ \bibnamefont
  {Sebastian}},\ }\href {\doibase 10.1103/PhysRevE.98.052138} {\bibfield
  {journal} {\bibinfo  {journal} {Phys. Rev. E}\ }\textbf {\bibinfo {volume}
  {98}},\ \bibinfo {pages} {052138} (\bibinfo {year} {2018})}\BibitemShut
  {NoStop}%
\bibitem [{\citenamefont {Sabri}\ \emph {et~al.}(2020)\citenamefont {Sabri},
  \citenamefont {Xu}, \citenamefont {Krapf},\ and\ \citenamefont
  {Weiss}}]{SabriXuKrapfWeiss2020}%
  \BibitemOpen
  \bibfield  {author} {\bibinfo {author} {\bibfnamefont {A.}~\bibnamefont
  {Sabri}}, \bibinfo {author} {\bibfnamefont {X.}~\bibnamefont {Xu}}, \bibinfo
  {author} {\bibfnamefont {D.}~\bibnamefont {Krapf}}, \ and\ \bibinfo {author}
  {\bibfnamefont {M.}~\bibnamefont {Weiss}},\ }\href {\doibase
  10.1103/PhysRevLett.125.058101} {\bibfield  {journal} {\bibinfo  {journal}
  {Phys. Rev. Lett.}\ }\textbf {\bibinfo {volume} {125}},\ \bibinfo {pages}
  {058101} (\bibinfo {year} {2020})}\BibitemShut {NoStop}%
\bibitem [{\citenamefont {Hidalgo-Soria}\ and\ \citenamefont
  {Barkai}(2020)}]{Hidalgo-SoriaBarkai2020}%
  \BibitemOpen
  \bibfield  {author} {\bibinfo {author} {\bibfnamefont {M.}~\bibnamefont
  {Hidalgo-Soria}}\ and\ \bibinfo {author} {\bibfnamefont {E.}~\bibnamefont
  {Barkai}},\ }\href {\doibase 10.1103/PhysRevE.102.012109} {\bibfield
  {journal} {\bibinfo  {journal} {Phys. Rev. E}\ }\textbf {\bibinfo {volume}
  {102}},\ \bibinfo {pages} {012109} (\bibinfo {year} {2020})}\BibitemShut
  {NoStop}%
\bibitem [{\citenamefont {Sposini}\ \emph {et~al.}(2020)\citenamefont
  {Sposini}, \citenamefont {Grebenkov}, \citenamefont {Metzler}, \citenamefont
  {Oshanin},\ and\ \citenamefont
  {Seno}}]{SposiniGrebenkovMetzlerOshaninSeno2020}%
  \BibitemOpen
  \bibfield  {author} {\bibinfo {author} {\bibfnamefont {V.}~\bibnamefont
  {Sposini}}, \bibinfo {author} {\bibfnamefont {D.}~\bibnamefont {Grebenkov}},
  \bibinfo {author} {\bibfnamefont {R.}~\bibnamefont {Metzler}}, \bibinfo
  {author} {\bibfnamefont {G.}~\bibnamefont {Oshanin}}, \ and\ \bibinfo
  {author} {\bibfnamefont {F.}~\bibnamefont {Seno}},\ }\href@noop {} {\bibfield
   {journal} {\bibinfo  {journal} {New J. Phys.}\ }\textbf {\bibinfo {volume}
  {22}},\ \bibinfo {pages} {063056} (\bibinfo {year} {2020})}\BibitemShut
  {NoStop}%
\bibitem [{\citenamefont {Barkai}\ and\ \citenamefont
  {Burov}(2020)}]{BarkaiBurov2020}%
  \BibitemOpen
  \bibfield  {author} {\bibinfo {author} {\bibfnamefont {E.}~\bibnamefont
  {Barkai}}\ and\ \bibinfo {author} {\bibfnamefont {S.}~\bibnamefont {Burov}},\
  }\href {\doibase 10.1103/PhysRevLett.124.060603} {\bibfield  {journal}
  {\bibinfo  {journal} {Phys. Rev. Lett.}\ }\textbf {\bibinfo {volume} {124}},\
  \bibinfo {pages} {060603} (\bibinfo {year} {2020})}\BibitemShut {NoStop}%
\bibitem [{\citenamefont {Wang}\ \emph {et~al.}(2020)\citenamefont {Wang},
  \citenamefont {Seno}, \citenamefont {Sokolov}, \citenamefont {Chechkin},\
  and\ \citenamefont {Metzler}}]{WangSenoSokolovChechkinMetzler2020}%
  \BibitemOpen
  \bibfield  {author} {\bibinfo {author} {\bibfnamefont {W.}~\bibnamefont
  {Wang}}, \bibinfo {author} {\bibfnamefont {F.}~\bibnamefont {Seno}}, \bibinfo
  {author} {\bibfnamefont {I.~M.}\ \bibnamefont {Sokolov}}, \bibinfo {author}
  {\bibfnamefont {A.~V.}\ \bibnamefont {Chechkin}}, \ and\ \bibinfo {author}
  {\bibfnamefont {R.}~\bibnamefont {Metzler}},\ }\href {\doibase DOI:
  10.1088/1367-2630/aba390} {\bibfield  {journal} {\bibinfo  {journal} {New. J.
  Phys.}\ }\textbf {\bibinfo {volume} {22}},\ \bibinfo {pages} {083041}
  (\bibinfo {year} {2020})}\BibitemShut {NoStop}%
\bibitem [{\citenamefont {Miyaguchi}(2017)}]{Miyaguchi2017}%
  \BibitemOpen
  \bibfield  {author} {\bibinfo {author} {\bibfnamefont {T.}~\bibnamefont
  {Miyaguchi}},\ }\href {\doibase 10.1103/PhysRevE.96.042501} {\bibfield
  {journal} {\bibinfo  {journal} {Phys. Rev. E}\ }\textbf {\bibinfo {volume}
  {96}},\ \bibinfo {pages} {042501} (\bibinfo {year} {2017})}\BibitemShut
  {NoStop}%
\bibitem [{\citenamefont {Honda}\ \emph {et~al.}(2008)\citenamefont {Honda},
  \citenamefont {Akiba}, \citenamefont {Kato}, \citenamefont {Sawada},
  \citenamefont {Sekijima}, \citenamefont {Ishimura}, \citenamefont {Ooishi},
  \citenamefont {Watanabe}, \citenamefont {Odahara},\ and\ \citenamefont
  {Harata}}]{HondaAkibaKatoSawadaSekijimaIshimuraOoishiWatanabeOdaharaHarata2008}%
  \BibitemOpen
  \bibfield  {author} {\bibinfo {author} {\bibfnamefont {S.}~\bibnamefont
  {Honda}}, \bibinfo {author} {\bibfnamefont {T.}~\bibnamefont {Akiba}},
  \bibinfo {author} {\bibfnamefont {Y.~S.}\ \bibnamefont {Kato}}, \bibinfo
  {author} {\bibfnamefont {Y.}~\bibnamefont {Sawada}}, \bibinfo {author}
  {\bibfnamefont {M.}~\bibnamefont {Sekijima}}, \bibinfo {author}
  {\bibfnamefont {M.}~\bibnamefont {Ishimura}}, \bibinfo {author}
  {\bibfnamefont {A.}~\bibnamefont {Ooishi}}, \bibinfo {author} {\bibfnamefont
  {H.}~\bibnamefont {Watanabe}}, \bibinfo {author} {\bibfnamefont
  {T.}~\bibnamefont {Odahara}}, \ and\ \bibinfo {author} {\bibfnamefont
  {K.}~\bibnamefont {Harata}},\ }\href@noop {} {\bibfield  {journal} {\bibinfo
  {journal} {J. Am. Chem. Soc.}\ }\textbf {\bibinfo {volume} {130}},\ \bibinfo
  {pages} {15327} (\bibinfo {year} {2008})}\BibitemShut {NoStop}%
\bibitem [{sup(ures)}]{support}%
  \BibitemOpen
  \href@noop {} {See Supplemental Material for details of MD simulations, analysis, and additional figures, which includes Refs. [38-50]}\BibitemShut {NoStop}%
\bibitem [{\citenamefont {Kubelka}\ \emph {et~al.}(2006)\citenamefont
  {Kubelka}, \citenamefont {Chiu}, \citenamefont {Davies}, \citenamefont
  {Eaton},\ and\ \citenamefont
  {Hofrichter}}]{KubelkaChiuDaviesEatonHofrichter2006}%
  \BibitemOpen
  \bibfield  {author} {\bibinfo {author} {\bibfnamefont {J.}~\bibnamefont
  {Kubelka}}, \bibinfo {author} {\bibfnamefont {T.~K.}\ \bibnamefont {Chiu}},
  \bibinfo {author} {\bibfnamefont {D.~R.}\ \bibnamefont {Davies}}, \bibinfo
  {author} {\bibfnamefont {W.~A.}\ \bibnamefont {Eaton}}, \ and\ \bibinfo
  {author} {\bibfnamefont {J.}~\bibnamefont {Hofrichter}},\ }\href@noop {}
  {\bibfield  {journal} {\bibinfo  {journal} {J. Mol. Biol.}\ }\textbf
  {\bibinfo {volume} {359}},\ \bibinfo {pages} {546} (\bibinfo {year}
  {2006})}\BibitemShut {NoStop}%
\bibitem [{\citenamefont {Ranganathan}\ \emph {et~al.}(1997)\citenamefont
  {Ranganathan}, \citenamefont {Lu}, \citenamefont {Hunter},\ and\
  \citenamefont {Noel}}]{RanganathanLuHunterNoel1997}%
  \BibitemOpen
  \bibfield  {author} {\bibinfo {author} {\bibfnamefont {R.}~\bibnamefont
  {Ranganathan}}, \bibinfo {author} {\bibfnamefont {K.~P.}\ \bibnamefont {Lu}},
  \bibinfo {author} {\bibfnamefont {T.}~\bibnamefont {Hunter}}, \ and\ \bibinfo
  {author} {\bibfnamefont {J.~P.}\ \bibnamefont {Noel}},\ }\href@noop {}
  {\bibfield  {journal} {\bibinfo  {journal} {Cell}\ }\textbf {\bibinfo
  {volume} {89}},\ \bibinfo {pages} {875} (\bibinfo {year} {1997})}\BibitemShut
  {NoStop}%
\bibitem [{\citenamefont {Abraham}\ \emph {et~al.}(2015)\citenamefont
  {Abraham}, \citenamefont {Murtola}, \citenamefont {Schulz}, \citenamefont
  {P{\'a}ll}, \citenamefont {Smith}, \citenamefont {Hess},\ and\ \citenamefont
  {Lindahl}}]{AbrahamMurtolaSchulzPallSmithHessLindahl2015}%
  \BibitemOpen
  \bibfield  {author} {\bibinfo {author} {\bibfnamefont {M.~J.}\ \bibnamefont
  {Abraham}}, \bibinfo {author} {\bibfnamefont {T.}~\bibnamefont {Murtola}},
  \bibinfo {author} {\bibfnamefont {R.}~\bibnamefont {Schulz}}, \bibinfo
  {author} {\bibfnamefont {S.}~\bibnamefont {P{\'a}ll}}, \bibinfo {author}
  {\bibfnamefont {J.~C.}\ \bibnamefont {Smith}}, \bibinfo {author}
  {\bibfnamefont {B.}~\bibnamefont {Hess}}, \ and\ \bibinfo {author}
  {\bibfnamefont {E.}~\bibnamefont {Lindahl}},\ }\href@noop {} {\bibfield
  {journal} {\bibinfo  {journal} {SoftwareX}\ }\textbf {\bibinfo {volume}
  {1}},\ \bibinfo {pages} {19} (\bibinfo {year} {2015})}\BibitemShut {NoStop}%
\bibitem [{\citenamefont {Okumura}(2012)}]{Okumura2012}%
  \BibitemOpen
  \bibfield  {author} {\bibinfo {author} {\bibfnamefont {H.}~\bibnamefont
  {Okumura}},\ }\href@noop {} {\bibfield  {journal} {\bibinfo  {journal}
  {Proteins}\ }\textbf {\bibinfo {volume} {80}},\ \bibinfo {pages} {2397}
  (\bibinfo {year} {2012})}\BibitemShut {NoStop}%
\bibitem [{\citenamefont {Berendsen}\ \emph {et~al.}(1984)\citenamefont
  {Berendsen}, \citenamefont {Postma}, \citenamefont {van Gunsteren},
  \citenamefont {DiNola},\ and\ \citenamefont
  {Haak}}]{BerendsenPostmaGunsterenDiNolaHaak1984}%
  \BibitemOpen
  \bibfield  {author} {\bibinfo {author} {\bibfnamefont {H.~J.~C.}\
  \bibnamefont {Berendsen}}, \bibinfo {author} {\bibfnamefont {J.~P.~M.}\
  \bibnamefont {Postma}}, \bibinfo {author} {\bibfnamefont {W.~F.}\
  \bibnamefont {van Gunsteren}}, \bibinfo {author} {\bibfnamefont
  {A.}~\bibnamefont {DiNola}}, \ and\ \bibinfo {author} {\bibfnamefont {J.~R.}\
  \bibnamefont {Haak}},\ }\href@noop {} {\bibfield  {journal} {\bibinfo
  {journal} {J. Chem. Phys.}\ }\textbf {\bibinfo {volume} {81}},\ \bibinfo
  {pages} {3684} (\bibinfo {year} {1984})}\BibitemShut {NoStop}%
\bibitem [{\citenamefont {Bussi}\ \emph {et~al.}(2009)\citenamefont {Bussi},
  \citenamefont {Zykova-Timan},\ and\ \citenamefont
  {Parrinello}}]{BussiZykova-TimanParrinello2009}%
  \BibitemOpen
  \bibfield  {author} {\bibinfo {author} {\bibfnamefont {G.}~\bibnamefont
  {Bussi}}, \bibinfo {author} {\bibfnamefont {T.}~\bibnamefont {Zykova-Timan}},
  \ and\ \bibinfo {author} {\bibfnamefont {M.}~\bibnamefont {Parrinello}},\
  }\href@noop {} {\bibfield  {journal} {\bibinfo  {journal} {J. Chem. Phys.}\
  }\textbf {\bibinfo {volume} {130}},\ \bibinfo {pages} {074101} (\bibinfo
  {year} {2009})}\BibitemShut {NoStop}%
\bibitem [{\citenamefont {Lindorff-Larsen}\ \emph {et~al.}(2010)\citenamefont
  {Lindorff-Larsen}, \citenamefont {Piana}, \citenamefont {Palmo},
  \citenamefont {Maragakis}, \citenamefont {Klepeis}, \citenamefont {Dror},\
  and\ \citenamefont
  {Shaw}}]{Lindorff-LarsenPianaPalmoMaragakisKlepeisDrorShaw2010}%
  \BibitemOpen
  \bibfield  {author} {\bibinfo {author} {\bibfnamefont {K.}~\bibnamefont
  {Lindorff-Larsen}}, \bibinfo {author} {\bibfnamefont {S.}~\bibnamefont
  {Piana}}, \bibinfo {author} {\bibfnamefont {K.}~\bibnamefont {Palmo}},
  \bibinfo {author} {\bibfnamefont {P.}~\bibnamefont {Maragakis}}, \bibinfo
  {author} {\bibfnamefont {J.~L.}\ \bibnamefont {Klepeis}}, \bibinfo {author}
  {\bibfnamefont {R.~O.}\ \bibnamefont {Dror}}, \ and\ \bibinfo {author}
  {\bibfnamefont {D.~E.}\ \bibnamefont {Shaw}},\ }\href@noop {} {\bibfield
  {journal} {\bibinfo  {journal} {Proteins}\ }\textbf {\bibinfo {volume}
  {78}},\ \bibinfo {pages} {1950} (\bibinfo {year} {2010})}\BibitemShut
  {NoStop}%
\bibitem [{\citenamefont {Jorgensen}\ \emph {et~al.}(1983)\citenamefont
  {Jorgensen}, \citenamefont {Chandrasekhar}, \citenamefont {Madura},
  \citenamefont {Impey},\ and\ \citenamefont
  {Klein}}]{JorgensenChandrasekharMaduraImpeyKlein1983}%
  \BibitemOpen
  \bibfield  {author} {\bibinfo {author} {\bibfnamefont {W.~L.}\ \bibnamefont
  {Jorgensen}}, \bibinfo {author} {\bibfnamefont {J.}~\bibnamefont
  {Chandrasekhar}}, \bibinfo {author} {\bibfnamefont {J.~D.}\ \bibnamefont
  {Madura}}, \bibinfo {author} {\bibfnamefont {R.~W.}\ \bibnamefont {Impey}}, \
  and\ \bibinfo {author} {\bibfnamefont {M.~L.}\ \bibnamefont {Klein}},\
  }\href@noop {} {\bibfield  {journal} {\bibinfo  {journal} {J. Chem. Phys.}\
  }\textbf {\bibinfo {volume} {79}},\ \bibinfo {pages} {926} (\bibinfo {year}
  {1983})}\BibitemShut {NoStop}%
\bibitem [{\citenamefont {Hess}\ \emph {et~al.}(1997)\citenamefont {Hess},
  \citenamefont {Bekker}, \citenamefont {Berendsen},\ and\ \citenamefont
  {Fraaije}}]{HessBekkerBerendsenFraaije1997}%
  \BibitemOpen
  \bibfield  {author} {\bibinfo {author} {\bibfnamefont {B.}~\bibnamefont
  {Hess}}, \bibinfo {author} {\bibfnamefont {H.}~\bibnamefont {Bekker}},
  \bibinfo {author} {\bibfnamefont {H.~J.~C.}\ \bibnamefont {Berendsen}}, \
  and\ \bibinfo {author} {\bibfnamefont {J.~G. E.~M.}\ \bibnamefont
  {Fraaije}},\ }\href@noop {} {\bibfield  {journal} {\bibinfo  {journal} {J.
  Comput. Chem.}\ }\textbf {\bibinfo {volume} {18}},\ \bibinfo {pages} {1463}
  (\bibinfo {year} {1997})}\BibitemShut {NoStop}%
\bibitem [{\citenamefont {Essmann}\ \emph {et~al.}(1995)\citenamefont
  {Essmann}, \citenamefont {Perera}, \citenamefont {Berkowitz}, \citenamefont
  {Darden}, \citenamefont {Lee},\ and\ \citenamefont
  {Pedersen}}]{EssmannPereraBerkowitzDardenLeePedersen1995}%
  \BibitemOpen
  \bibfield  {author} {\bibinfo {author} {\bibfnamefont {U.}~\bibnamefont
  {Essmann}}, \bibinfo {author} {\bibfnamefont {L.}~\bibnamefont {Perera}},
  \bibinfo {author} {\bibfnamefont {M.~L.}\ \bibnamefont {Berkowitz}}, \bibinfo
  {author} {\bibfnamefont {T.}~\bibnamefont {Darden}}, \bibinfo {author}
  {\bibfnamefont {H.}~\bibnamefont {Lee}}, \ and\ \bibinfo {author}
  {\bibfnamefont {L.~G.}\ \bibnamefont {Pedersen}},\ }\href@noop {} {\bibfield
  {journal} {\bibinfo  {journal} {J. Chem. Phys.}\ }\textbf {\bibinfo {volume}
  {103}},\ \bibinfo {pages} {8577} (\bibinfo {year} {1995})}\BibitemShut
  {NoStop}%
\bibitem [{\citenamefont {Mitsutake}\ and\ \citenamefont
  {Takano}(2018)}]{MitsutakeTakano2018}%
  \BibitemOpen
  \bibfield  {author} {\bibinfo {author} {\bibfnamefont {A.}~\bibnamefont
  {Mitsutake}}\ and\ \bibinfo {author} {\bibfnamefont {H.}~\bibnamefont
  {Takano}},\ }\href@noop {} {\bibfield  {journal} {\bibinfo  {journal}
  {Biophys. Rev.}\ }\textbf {\bibinfo {volume} {10}},\ \bibinfo {pages} {375}
  (\bibinfo {year} {2018})}\BibitemShut {NoStop}%
\bibitem [{\citenamefont {Naritomi}\ and\ \citenamefont
  {Fuchigami}(2011)}]{NaritomiFuchigami2011}%
  \BibitemOpen
  \bibfield  {author} {\bibinfo {author} {\bibfnamefont {Y.}~\bibnamefont
  {Naritomi}}\ and\ \bibinfo {author} {\bibfnamefont {S.}~\bibnamefont
  {Fuchigami}},\ }\href@noop {} {\bibfield  {journal} {\bibinfo  {journal} {J.
  Chem. Phys.}\ }\textbf {\bibinfo {volume} {134}},\ \bibinfo {pages} {02B617}
  (\bibinfo {year} {2011})}\BibitemShut {NoStop}%
\bibitem [{\citenamefont {Durchschlag}\ and\ \citenamefont
  {Zipper}(1997)}]{DurchschlagZipper1997}%
  \BibitemOpen
  \bibfield  {author} {\bibinfo {author} {\bibfnamefont {H.}~\bibnamefont
  {Durchschlag}}\ and\ \bibinfo {author} {\bibfnamefont {P.}~\bibnamefont
  {Zipper}},\ }\href@noop {} {\bibfield  {journal} {\bibinfo  {journal} {J.
  Appl. Crystallogr.}\ }\textbf {\bibinfo {volume} {30}},\ \bibinfo {pages}
  {1112} (\bibinfo {year} {1997})}\BibitemShut {NoStop}%
\bibitem [{\citenamefont {Iben}\ \emph {et~al.}(1989)\citenamefont {Iben},
  \citenamefont {Braunstein}, \citenamefont {Doster}, \citenamefont
  {Frauenfelder}, \citenamefont {Hong}, \citenamefont {Johnson}, \citenamefont
  {Luck}, \citenamefont {Ormos}, \citenamefont {Schulte}, \citenamefont
  {Steinbach}, \citenamefont {Xie},\ and\ \citenamefont
  {Young}}]{IbenBraunsteinDosterFrauenfelderHongJohnsonLuckOrmosSchulteSteinbachXieYoung1989}%
  \BibitemOpen
  \bibfield  {author} {\bibinfo {author} {\bibfnamefont {I.~E.~T.}\
  \bibnamefont {Iben}}, \bibinfo {author} {\bibfnamefont {D.}~\bibnamefont
  {Braunstein}}, \bibinfo {author} {\bibfnamefont {W.}~\bibnamefont {Doster}},
  \bibinfo {author} {\bibfnamefont {H.}~\bibnamefont {Frauenfelder}}, \bibinfo
  {author} {\bibfnamefont {M.~K.}\ \bibnamefont {Hong}}, \bibinfo {author}
  {\bibfnamefont {J.~B.}\ \bibnamefont {Johnson}}, \bibinfo {author}
  {\bibfnamefont {S.}~\bibnamefont {Luck}}, \bibinfo {author} {\bibfnamefont
  {P.}~\bibnamefont {Ormos}}, \bibinfo {author} {\bibfnamefont
  {A.}~\bibnamefont {Schulte}}, \bibinfo {author} {\bibfnamefont {P.~J.}\
  \bibnamefont {Steinbach}}, \bibinfo {author} {\bibfnamefont {A.~H.}\
  \bibnamefont {Xie}}, \ and\ \bibinfo {author} {\bibfnamefont {R.~D.}\
  \bibnamefont {Young}},\ }\href {\doibase 10.1103/PhysRevLett.62.1916}
  {\bibfield  {journal} {\bibinfo  {journal} {Phys. Rev. Lett.}\ }\textbf
  {\bibinfo {volume} {62}},\ \bibinfo {pages} {1916} (\bibinfo {year}
  {1989})}\BibitemShut {NoStop}%
\bibitem [{\citenamefont {Takano}\ \emph {et~al.}(1998)\citenamefont {Takano},
  \citenamefont {Takahashi},\ and\ \citenamefont
  {Nagayama}}]{TakanoTakahashiNagayama1998}%
  \BibitemOpen
  \bibfield  {author} {\bibinfo {author} {\bibfnamefont {M.}~\bibnamefont
  {Takano}}, \bibinfo {author} {\bibfnamefont {T.}~\bibnamefont {Takahashi}}, \
  and\ \bibinfo {author} {\bibfnamefont {K.}~\bibnamefont {Nagayama}},\ }\href
  {\doibase 10.1103/PhysRevLett.80.5691} {\bibfield  {journal} {\bibinfo
  {journal} {Phys. Rev. Lett.}\ }\textbf {\bibinfo {volume} {80}},\ \bibinfo
  {pages} {5691} (\bibinfo {year} {1998})}\BibitemShut {NoStop}%
\bibitem [{\citenamefont {Yang}\ \emph {et~al.}(2003)\citenamefont {Yang},
  \citenamefont {Luo}, \citenamefont {Karnchanaphanurach}, \citenamefont
  {Louie}, \citenamefont {Rech}, \citenamefont {Cova}, \citenamefont {Xun},\
  and\ \citenamefont {Xie}}]{YangLuoKarnchanaphanurachLouieRechCovaXunXie2003}%
  \BibitemOpen
  \bibfield  {author} {\bibinfo {author} {\bibfnamefont {H.}~\bibnamefont
  {Yang}}, \bibinfo {author} {\bibfnamefont {G.}~\bibnamefont {Luo}}, \bibinfo
  {author} {\bibfnamefont {P.}~\bibnamefont {Karnchanaphanurach}}, \bibinfo
  {author} {\bibfnamefont {T.~M.}\ \bibnamefont {Louie}}, \bibinfo {author}
  {\bibfnamefont {I.}~\bibnamefont {Rech}}, \bibinfo {author} {\bibfnamefont
  {S.}~\bibnamefont {Cova}}, \bibinfo {author} {\bibfnamefont {L.}~\bibnamefont
  {Xun}}, \ and\ \bibinfo {author} {\bibfnamefont {X.~S.}\ \bibnamefont
  {Xie}},\ }\href@noop {} {\bibfield  {journal} {\bibinfo  {journal} {Science}\
  }\textbf {\bibinfo {volume} {302}},\ \bibinfo {pages} {262} (\bibinfo {year}
  {2003})}\BibitemShut {NoStop}%
\bibitem [{\citenamefont {Yamamoto}\ \emph
  {et~al.}(2014{\natexlab{a}})\citenamefont {Yamamoto}, \citenamefont
  {Akimoto}, \citenamefont {Hirano}, \citenamefont {Yasui},\ and\ \citenamefont
  {Yasuoka}}]{YamamotoAkimotoHiranoYasuiYasuoka2014}%
  \BibitemOpen
  \bibfield  {author} {\bibinfo {author} {\bibfnamefont {E.}~\bibnamefont
  {Yamamoto}}, \bibinfo {author} {\bibfnamefont {T.}~\bibnamefont {Akimoto}},
  \bibinfo {author} {\bibfnamefont {Y.}~\bibnamefont {Hirano}}, \bibinfo
  {author} {\bibfnamefont {M.}~\bibnamefont {Yasui}}, \ and\ \bibinfo {author}
  {\bibfnamefont {K.}~\bibnamefont {Yasuoka}},\ }\href {\doibase
  10.1103/PhysRevE.89.022718} {\bibfield  {journal} {\bibinfo  {journal} {Phys.
  Rev. E}\ }\textbf {\bibinfo {volume} {89}},\ \bibinfo {pages} {022718}
  (\bibinfo {year} {2014}{\natexlab{a}})}\BibitemShut {NoStop}%
\bibitem [{\citenamefont {Niemann}\ \emph {et~al.}(2013)\citenamefont
  {Niemann}, \citenamefont {Kantz},\ and\ \citenamefont
  {Barkai}}]{NiemannKantzBarkai2013}%
  \BibitemOpen
  \bibfield  {author} {\bibinfo {author} {\bibfnamefont {M.}~\bibnamefont
  {Niemann}}, \bibinfo {author} {\bibfnamefont {H.}~\bibnamefont {Kantz}}, \
  and\ \bibinfo {author} {\bibfnamefont {E.}~\bibnamefont {Barkai}},\ }\href
  {\doibase 10.1103/PhysRevLett.110.140603} {\bibfield  {journal} {\bibinfo
  {journal} {Phys. Rev. Lett.}\ }\textbf {\bibinfo {volume} {110}},\ \bibinfo
  {pages} {140603} (\bibinfo {year} {2013})}\BibitemShut {NoStop}%
\bibitem [{\citenamefont {Sadegh}\ \emph {et~al.}(2014)\citenamefont {Sadegh},
  \citenamefont {Barkai},\ and\ \citenamefont {Krapf}}]{SadeghBarkaiKrapf2014}%
  \BibitemOpen
  \bibfield  {author} {\bibinfo {author} {\bibfnamefont {S.}~\bibnamefont
  {Sadegh}}, \bibinfo {author} {\bibfnamefont {E.}~\bibnamefont {Barkai}}, \
  and\ \bibinfo {author} {\bibfnamefont {D.}~\bibnamefont {Krapf}},\
  }\href@noop {} {\bibfield  {journal} {\bibinfo  {journal} {New J. Phys.}\
  }\textbf {\bibinfo {volume} {16}},\ \bibinfo {pages} {113054} (\bibinfo
  {year} {2014})}\BibitemShut {NoStop}%
\bibitem [{\citenamefont {Leibovich}\ and\ \citenamefont
  {Barkai}(2015)}]{LeibovichBarkai2015}%
  \BibitemOpen
  \bibfield  {author} {\bibinfo {author} {\bibfnamefont {N.}~\bibnamefont
  {Leibovich}}\ and\ \bibinfo {author} {\bibfnamefont {E.}~\bibnamefont
  {Barkai}},\ }\href {\doibase 10.1103/PhysRevLett.115.080602} {\bibfield
  {journal} {\bibinfo  {journal} {Phys. Rev. Lett.}\ }\textbf {\bibinfo
  {volume} {115}},\ \bibinfo {pages} {080602} (\bibinfo {year}
  {2015})}\BibitemShut {NoStop}%
\bibitem [{\citenamefont {Takano}\ and\ \citenamefont
  {Miyashita}(1995)}]{TakanoMiyashita1995}%
  \BibitemOpen
  \bibfield  {author} {\bibinfo {author} {\bibfnamefont {H.}~\bibnamefont
  {Takano}}\ and\ \bibinfo {author} {\bibfnamefont {S.}~\bibnamefont
  {Miyashita}},\ }\href@noop {} {\bibfield  {journal} {\bibinfo  {journal} {J.
  Phys. Soc. Jpn.}\ }\textbf {\bibinfo {volume} {64}},\ \bibinfo {pages} {3688}
  (\bibinfo {year} {1995})}\BibitemShut {NoStop}%
\bibitem [{\citenamefont {Hirao}\ \emph {et~al.}(1997)\citenamefont {Hirao},
  \citenamefont {Koseki},\ and\ \citenamefont
  {Takano}}]{HiraoKosekiTakano1997}%
  \BibitemOpen
  \bibfield  {author} {\bibinfo {author} {\bibfnamefont {H.}~\bibnamefont
  {Hirao}}, \bibinfo {author} {\bibfnamefont {S.}~\bibnamefont {Koseki}}, \
  and\ \bibinfo {author} {\bibfnamefont {H.}~\bibnamefont {Takano}},\
  }\href@noop {} {\bibfield  {journal} {\bibinfo  {journal} {J. Phys. Soc.
  Jpn.}\ }\textbf {\bibinfo {volume} {66}},\ \bibinfo {pages} {3399} (\bibinfo
  {year} {1997})}\BibitemShut {NoStop}%
\bibitem [{\citenamefont {Mitsutake}\ \emph {et~al.}(2011)\citenamefont
  {Mitsutake}, \citenamefont {Iijima},\ and\ \citenamefont
  {Takano}}]{MitsutakeIijimaTakano2011}%
  \BibitemOpen
  \bibfield  {author} {\bibinfo {author} {\bibfnamefont {A.}~\bibnamefont
  {Mitsutake}}, \bibinfo {author} {\bibfnamefont {H.}~\bibnamefont {Iijima}}, \
  and\ \bibinfo {author} {\bibfnamefont {H.}~\bibnamefont {Takano}},\
  }\href@noop {} {\bibfield  {journal} {\bibinfo  {journal} {J. Chem. Phys.}\
  }\textbf {\bibinfo {volume} {135}},\ \bibinfo {pages} {164102} (\bibinfo
  {year} {2011})}\BibitemShut {NoStop}%
\bibitem [{\citenamefont {Mitsutake}\ and\ \citenamefont
  {Takano}(2015)}]{MitsutakeTakano2015}%
  \BibitemOpen
  \bibfield  {author} {\bibinfo {author} {\bibfnamefont {A.}~\bibnamefont
  {Mitsutake}}\ and\ \bibinfo {author} {\bibfnamefont {H.}~\bibnamefont
  {Takano}},\ }\href@noop {} {\bibfield  {journal} {\bibinfo  {journal} {J.
  Chem. Phys.}\ }\textbf {\bibinfo {volume} {143}},\ \bibinfo {pages} {124111}
  (\bibinfo {year} {2015})}\BibitemShut {NoStop}%
\bibitem [{\citenamefont {He}\ \emph {et~al.}(2008)\citenamefont {He},
  \citenamefont {Burov}, \citenamefont {Metzler},\ and\ \citenamefont
  {Barkai}}]{HeBurovMetzlerBarkai2008}%
  \BibitemOpen
  \bibfield  {author} {\bibinfo {author} {\bibfnamefont {Y.}~\bibnamefont
  {He}}, \bibinfo {author} {\bibfnamefont {S.}~\bibnamefont {Burov}}, \bibinfo
  {author} {\bibfnamefont {R.}~\bibnamefont {Metzler}}, \ and\ \bibinfo
  {author} {\bibfnamefont {E.}~\bibnamefont {Barkai}},\ }\href {\doibase
  10.1103/PhysRevLett.101.058101} {\bibfield  {journal} {\bibinfo  {journal}
  {Phys. Rev. Lett.}\ }\textbf {\bibinfo {volume} {101}},\ \bibinfo {pages}
  {058101} (\bibinfo {year} {2008})}\BibitemShut {NoStop}%
\bibitem [{\citenamefont {Metzler}\ \emph {et~al.}(2014)\citenamefont
  {Metzler}, \citenamefont {Jeon}, \citenamefont {Cherstvy},\ and\
  \citenamefont {Barkai}}]{MetzlerJeonCherstvyBarkai2014}%
  \BibitemOpen
  \bibfield  {author} {\bibinfo {author} {\bibfnamefont {R.}~\bibnamefont
  {Metzler}}, \bibinfo {author} {\bibfnamefont {J.-H.}\ \bibnamefont {Jeon}},
  \bibinfo {author} {\bibfnamefont {A.~G.}\ \bibnamefont {Cherstvy}}, \ and\
  \bibinfo {author} {\bibfnamefont {E.}~\bibnamefont {Barkai}},\ }\href@noop {}
  {\bibfield  {journal} {\bibinfo  {journal} {Phys. Chem. Chem. Phys.}\
  }\textbf {\bibinfo {volume} {16}},\ \bibinfo {pages} {24128} (\bibinfo {year}
  {2014})}\BibitemShut {NoStop}%
\bibitem [{\citenamefont {Miyaguchi}\ and\ \citenamefont
  {Akimoto}(2011{\natexlab{a}})}]{MiyaguchiAkimoto2011a}%
  \BibitemOpen
  \bibfield  {author} {\bibinfo {author} {\bibfnamefont {T.}~\bibnamefont
  {Miyaguchi}}\ and\ \bibinfo {author} {\bibfnamefont {T.}~\bibnamefont
  {Akimoto}},\ }\href {\doibase 10.1103/PhysRevE.83.031926} {\bibfield
  {journal} {\bibinfo  {journal} {Phys. Rev. E}\ }\textbf {\bibinfo {volume}
  {83}},\ \bibinfo {pages} {031926} (\bibinfo {year}
  {2011}{\natexlab{a}})}\BibitemShut {NoStop}%
\bibitem [{\citenamefont {Miyaguchi}\ and\ \citenamefont
  {Akimoto}(2011{\natexlab{b}})}]{MiyaguchiAkimoto2011}%
  \BibitemOpen
  \bibfield  {author} {\bibinfo {author} {\bibfnamefont {T.}~\bibnamefont
  {Miyaguchi}}\ and\ \bibinfo {author} {\bibfnamefont {T.}~\bibnamefont
  {Akimoto}},\ }\href {\doibase 10.1103/PhysRevE.83.062101} {\bibfield
  {journal} {\bibinfo  {journal} {Phys. Rev. E}\ }\textbf {\bibinfo {volume}
  {83}},\ \bibinfo {pages} {062101} (\bibinfo {year}
  {2011}{\natexlab{b}})}\BibitemShut {NoStop}%
\bibitem [{\citenamefont {Akimoto}\ and\ \citenamefont
  {Yamamoto}(2016{\natexlab{b}})}]{AkimotoYamamoto2016a}%
  \BibitemOpen
  \bibfield  {author} {\bibinfo {author} {\bibfnamefont {T.}~\bibnamefont
  {Akimoto}}\ and\ \bibinfo {author} {\bibfnamefont {E.}~\bibnamefont
  {Yamamoto}},\ }\href@noop {} {\bibfield  {journal} {\bibinfo  {journal} {J.
  Stat. Mech.}\ }\textbf {\bibinfo {volume} {2016}},\ \bibinfo {pages} {123201}
  (\bibinfo {year} {2016}{\natexlab{b}})}\BibitemShut {NoStop}%
\bibitem [{\citenamefont {Zimm}(1956)}]{Zimm1956}%
  \BibitemOpen
  \bibfield  {author} {\bibinfo {author} {\bibfnamefont {B.~H.}\ \bibnamefont
  {Zimm}},\ }\href@noop {} {\bibfield  {journal} {\bibinfo  {journal} {J. Chem.
  Phys.}\ }\textbf {\bibinfo {volume} {24}},\ \bibinfo {pages} {269} (\bibinfo
  {year} {1956})}\BibitemShut {NoStop}%
\bibitem [{\citenamefont {Ermak}\ and\ \citenamefont
  {McCammon}(1978)}]{ErmakMcCammon1978}%
  \BibitemOpen
  \bibfield  {author} {\bibinfo {author} {\bibfnamefont {D.~L.}\ \bibnamefont
  {Ermak}}\ and\ \bibinfo {author} {\bibfnamefont {J.~A.}\ \bibnamefont
  {McCammon}},\ }\href@noop {} {\bibfield  {journal} {\bibinfo  {journal} {J.
  Chem. Phys.}\ }\textbf {\bibinfo {volume} {69}},\ \bibinfo {pages} {1352}
  (\bibinfo {year} {1978})}\BibitemShut {NoStop}%
\bibitem [{\citenamefont {Yamamoto}\ \emph
  {et~al.}(2014{\natexlab{b}})\citenamefont {Yamamoto}, \citenamefont
  {Akimoto}, \citenamefont {Yasui},\ and\ \citenamefont
  {Yasuoka}}]{YamamotoAkimotoYasuiYasuoka2014}%
  \BibitemOpen
  \bibfield  {author} {\bibinfo {author} {\bibfnamefont {E.}~\bibnamefont
  {Yamamoto}}, \bibinfo {author} {\bibfnamefont {T.}~\bibnamefont {Akimoto}},
  \bibinfo {author} {\bibfnamefont {M.}~\bibnamefont {Yasui}}, \ and\ \bibinfo
  {author} {\bibfnamefont {K.}~\bibnamefont {Yasuoka}},\ }\href@noop {}
  {\bibfield  {journal} {\bibinfo  {journal} {Sci. Rep.}\ }\textbf {\bibinfo
  {volume} {4}},\ \bibinfo {pages} {4720} (\bibinfo {year}
  {2014}{\natexlab{b}})}\BibitemShut {NoStop}%
\bibitem [{\citenamefont {Tan}\ \emph {et~al.}(2018)\citenamefont {Tan},
  \citenamefont {Liang}, \citenamefont {Xu}, \citenamefont {Mamontov},
  \citenamefont {Li}, \citenamefont {Xing},\ and\ \citenamefont
  {Hong}}]{TanLiangXuMamontovLiXingHong2018}%
  \BibitemOpen
  \bibfield  {author} {\bibinfo {author} {\bibfnamefont {P.}~\bibnamefont
  {Tan}}, \bibinfo {author} {\bibfnamefont {Y.}~\bibnamefont {Liang}}, \bibinfo
  {author} {\bibfnamefont {Q.}~\bibnamefont {Xu}}, \bibinfo {author}
  {\bibfnamefont {E.}~\bibnamefont {Mamontov}}, \bibinfo {author}
  {\bibfnamefont {J.}~\bibnamefont {Li}}, \bibinfo {author} {\bibfnamefont
  {X.}~\bibnamefont {Xing}}, \ and\ \bibinfo {author} {\bibfnamefont
  {L.}~\bibnamefont {Hong}},\ }\href {\doibase 10.1103/PhysRevLett.120.248101}
  {\bibfield  {journal} {\bibinfo  {journal} {Phys. Rev. Lett.}\ }\textbf
  {\bibinfo {volume} {120}},\ \bibinfo {pages} {248101} (\bibinfo {year}
  {2018})}\BibitemShut {NoStop}%
\bibitem [{\citenamefont {Krapf}\ and\ \citenamefont
  {Metzler}(2019)}]{KrapfMetzler2019}%
  \BibitemOpen
  \bibfield  {author} {\bibinfo {author} {\bibfnamefont {D.}~\bibnamefont
  {Krapf}}\ and\ \bibinfo {author} {\bibfnamefont {R.}~\bibnamefont
  {Metzler}},\ }\href@noop {} {\bibfield  {journal} {\bibinfo  {journal} {Phys.
  Today}\ }\textbf {\bibinfo {volume} {72}},\ \bibinfo {pages} {48} (\bibinfo
  {year} {2019})}\BibitemShut {NoStop}%
\bibitem [{\citenamefont {Akimoto}\ and\ \citenamefont
  {Yamamoto}(2017)}]{AkimotoYamamoto2017}%
  \BibitemOpen
  \bibfield  {author} {\bibinfo {author} {\bibfnamefont {T.}~\bibnamefont
  {Akimoto}}\ and\ \bibinfo {author} {\bibfnamefont {E.}~\bibnamefont
  {Yamamoto}},\ }\href {\doibase 10.1103/PhysRevE.96.052138} {\bibfield
  {journal} {\bibinfo  {journal} {Phys. Rev. E}\ }\textbf {\bibinfo {volume}
  {96}},\ \bibinfo {pages} {052138} (\bibinfo {year} {2017})}\BibitemShut
  {NoStop}%
\bibitem [{\citenamefont {Doi}\ and\ \citenamefont
  {Edwards}(1988)}]{DoiEdwards1988}%
  \BibitemOpen
  \bibfield  {author} {\bibinfo {author} {\bibfnamefont {M.}~\bibnamefont
  {Doi}}\ and\ \bibinfo {author} {\bibfnamefont {S.~F.}\ \bibnamefont
  {Edwards}},\ }\href@noop {} {\bibinfo  {journal} {The Theory of Polymer
  Dynamics}\ }(\bibinfo  {publisher} {Oxford University Press},\ \bibinfo
  {year} {1988})\BibitemShut {NoStop}%
\bibitem [{\citenamefont {Hu}\ \emph {et~al.}(2016)\citenamefont {Hu},
  \citenamefont {Hong}, \citenamefont {Smith}, \citenamefont {Neusius},
  \citenamefont {Cheng},\ and\ \citenamefont
  {Smith}}]{HuHongSmithNeusiusChengSmith2016}%
  \BibitemOpen
  \bibfield  {author} {\bibinfo {author} {\bibfnamefont {X.}~\bibnamefont
  {Hu}}, \bibinfo {author} {\bibfnamefont {L.}~\bibnamefont {Hong}}, \bibinfo
  {author} {\bibfnamefont {M.~D.}\ \bibnamefont {Smith}}, \bibinfo {author}
  {\bibfnamefont {T.}~\bibnamefont {Neusius}}, \bibinfo {author} {\bibfnamefont
  {X.}~\bibnamefont {Cheng}}, \ and\ \bibinfo {author} {\bibfnamefont {J.~C.}\
  \bibnamefont {Smith}},\ }\href@noop {} {\bibfield  {journal} {\bibinfo
  {journal} {Nat. Phys.}\ }\textbf {\bibinfo {volume} {12}},\ \bibinfo {pages}
  {171} (\bibinfo {year} {2016})}\BibitemShut {NoStop}%
\bibitem [{\citenamefont {Etoc}\ \emph {et~al.}(2018)\citenamefont {Etoc},
  \citenamefont {Balloul}, \citenamefont {Vicario}, \citenamefont {Normanno},
  \citenamefont {Li{\ss}e}, \citenamefont {Sittner}, \citenamefont {Piehler},
  \citenamefont {Dahan},\ and\ \citenamefont
  {Coppey}}]{EtocBalloulVicarioNormannoLiseSittnerPiehlerDahanCoppey2018}%
  \BibitemOpen
  \bibfield  {author} {\bibinfo {author} {\bibfnamefont {F.}~\bibnamefont
  {Etoc}}, \bibinfo {author} {\bibfnamefont {E.}~\bibnamefont {Balloul}},
  \bibinfo {author} {\bibfnamefont {C.}~\bibnamefont {Vicario}}, \bibinfo
  {author} {\bibfnamefont {D.}~\bibnamefont {Normanno}}, \bibinfo {author}
  {\bibfnamefont {D.}~\bibnamefont {Li{\ss}e}}, \bibinfo {author}
  {\bibfnamefont {A.}~\bibnamefont {Sittner}}, \bibinfo {author} {\bibfnamefont
  {J.}~\bibnamefont {Piehler}}, \bibinfo {author} {\bibfnamefont
  {M.}~\bibnamefont {Dahan}}, \ and\ \bibinfo {author} {\bibfnamefont
  {M.}~\bibnamefont {Coppey}},\ }\href@noop {} {\bibfield  {journal} {\bibinfo
  {journal} {Nat. Mater.}\ }\textbf {\bibinfo {volume} {17}},\ \bibinfo {pages}
  {740} (\bibinfo {year} {2018})}\BibitemShut {NoStop}%
\end{thebibliography}
\end{document}